\documentclass[11pt]{article}
\pdfoutput=1
\usepackage{verbatim}
\usepackage{amssymb}
\usepackage{xcolor}
\usepackage[centertags]{amsmath}
\usepackage[square, comma, sort&compress,numbers]{natbib}	
\usepackage{array,multirow}
\usepackage{graphicx}
\usepackage{bbold}
\usepackage{braket}
\usepackage{wrapfig}
\usepackage{float}
\usepackage{soul}
\usepackage{url}
\usepackage{esvect}
\numberwithin{equation}{section}

\usepackage{tikz,pgf}
\usetikzlibrary{shapes}
\usetikzlibrary{calc}
\usetikzlibrary{decorations.pathmorphing}
\usetikzlibrary{decorations.pathreplacing,shapes.misc}

\usetikzlibrary{positioning}
\usetikzlibrary{arrows}
\usetikzlibrary{decorations.markings}
\usetikzlibrary{shadings}
\usetikzlibrary{intersections}
\usepackage{jheppub}
\numberwithin{equation}{section} \makeatletter
\@addtoreset{equation}{section}

\newcommand{\be}{\begin{equation}}
\newcommand{\ee}{\end{equation}}

\newcommand{\dps}{\displaystyle}
\newcommand{\bL}{\mathbf{L}}
\newcommand{\bW}{\mathbf{W}}

\newcommand{\varW}{\mathcal{W}}

\newcommand{\mf}[1]{\mathfrak{#1}}
\newcommand{\mc}[1]{\mathcal{#1}}

\newcommand{\ocom}[1]{     #1, \bar{ #1 }}
\newcommand{\bref}[1]{\textbf{\ref{#1}}}

\renewcommand{\sl}{\mathfrak{sl}} 

\DeclareMathOperator{\Tr}{Tr}

\def\tild{  \tilde {\Delta}}
\def\dd2{ {\Delta, \bar{\Delta} }}
\def\dd3{{1-\Delta, 1-\bar{\Delta}}}

\makeatletter
\def\@fpheader{\vspace{-.1cm}}
\makeatother

\title{Global conformal blocks via Shadow formalism}

\author{Vladimir Belavin, J. Ramos Cabezas}

\affiliation{Physics Department, Ariel University, Ariel 40700, Israel.}

\emailAdd{vladimirbe@ariel.ac.il, juanjose.ramoscab@msmail.ariel.ac.il}

\abstract{We study $\mf{sl}_2$ and $\mf{sl}_3$ global conformal blocks on a sphere and a torus, using the shadow formalism. These blocks arise in the context of Virasoro and $\varW_3$ conformal field theories in the large central charge limit. In the $\mf{sl}_2$ case, we demonstrate that the shadow formalism yields the known expressions in terms of conformal partial waves. Then, we extend this approach to the $\mf{sl}_3$ case and show that it allows to build simple integral representations for $\mf{sl}_3$ global blocks. We demonstrate this construction on two examples: the four-point block on the sphere and the one-point torus block. } 

\begin{document}
\maketitle
\section{Introduction}

An important class of conformal field theories is the class CFTs possessing $\varW_N$ ($N\ge2$) symmetry, the generators of which are $N-1$ holomorphic fields of spin $2, ..., N$. The simplest case, $N=2$, corresponds to the conformal field theory with Virasoro symmetry, where the generator is the spin-2 energy-momentum tensor. Examples of conformal field theories with $\mathcal{W}_N$ symmetry include the $\mathfrak{sl}_N$ quantum Toda field theory, which is a generalization of Liouville field theory, corresponding to $N=2$ and arising in the description of two-dimensional quantum gravity. Thus, the $\mathfrak{sl}_N$ quantum Toda field theory is relevant for higher-spin generalizations of two-dimensional quantum gravity.

The correlation functions of primary fields in conformal field theories can generally be expressed in terms of model-independent building blocks, known as \textit{conformal blocks} (CBs). Apart from the direct application within the CFT frame, CBs play an essential role in different contexts, in particular they have dual interpretations in AdS/CFT correspondence, as geodesic networks, see, e.g., \cite{Hartman:2013mia, Fitzpatrick:2014vua, Caputa:2014eta, Hijano:2015rla, Fitzpatrick:2015zha, Alkalaev:2015wia, Hijano:2015qja, Hijano:2015zsa,  Alkalaev:2015lca, Alkalaev:2015fbw, Banerjee:2016qca, Gobeil:2018fzy, Hung:2018mcn, Alekseev:2019gkl, RamosCabezas:2020mew}, or specific Wilson line configurations, see, e.g., ~\cite{ deBoer:2013vca, Ammon:2013hba, deBoer:2014sna, Hegde:2015dqh, Melnikov:2016eun, Bhatta:2016hpz, Besken:2017fsj, Hikida:2017ehf, Hikida:2018eih, Hikida:2018dxe, Besken:2018zro, Bhatta:2018gjb, DHoker:2019clx, Castro:2018srf, Kraus:2018zrn, Hulik:2018dpl, Castro:2020smu, Chen:2020nlj,Belavin:2022bib, Alkalaev:2023axo}. CBs also appear in the context of solvable lattice models \cite{Fuchs:1993xu} through the connection of the braiding matrices with the Boltzmann weights of interaction-round-the-face lattice models (see, e.g.,~\cite{Gepner:1992kx, Fuchs:1993xu}).

In this work, we describe the CBs of the correlation functions of primary fields in conformal field theories with $\mathcal{W}_2$ (Virasoro) and $\mathcal{W}_3$ symmetries in the large central charge limit, using the so-called \textit{shadow formalism}~\cite{Ferrara:1972uq, Ferrara:1972kab, Simmons-Duffin:2012juh, Rosenhaus:2018zqn}. This method allows us to express the CBs in integral representations involving the so-called  \textit{conformal partial waves}. In the case of Virasoro CFT, this method has been well-studied in both spherical and torus topologies. Our goal is to generalize this method to the $\mathcal{W}_3$ CFT, particularly on the torus, where no exact expressions for the blocks are known.

The main object of our study is the one-point CB on the torus in the large central charge limit. The one-point correlation function of a primary field $\phi_{\Delta_1,\bar{\Delta}_1}$ with holomorphic and antiholomorphic conformal dimensions $\Delta_1, \bar{\Delta}_1$ on the torus is defined as\footnote{Throughout this paper, we omit the factor $(q\bar{q})^{-\frac{c}{24}}$, which can be easily restored.}
\begin{equation} 	 \label{main1ptblock}
\braket{\phi_{\Delta_1, \bar{\Delta}_1}(z_1, \bar{z}_1)}_{torus}= \Tr_{\tilde{\Delta} \in \mathcal{D}}\Big( q^{\bL_0} \bar{q}^{\bar{\bL}_0}\phi_{\Delta_1, \bar{\Delta}_1}(z_1, \bar{z}_1)    \Big)=\sum_{\tilde{\Delta} \in \mathcal{D}} C_{\tilde{\Delta} \Delta_1 \tilde{\Delta}} |\mathcal{F}(\Delta_1, \tilde{\Delta}, q)_A|^2 \; ,
\end{equation}
where $\Tr_{\tilde{\Delta}}$ denotes the trace taken over a module of the symmetry algebra $A$ associated with the primary field $\phi_{\tilde{\Delta}, \bar{\tilde{\Delta}}}$ in the intermediate OPE channel, $\mathcal{D}$ is the domain of primary fields of the corresponding conformal field theory, $q$ is the elliptic parameter of the torus $q=e^{2\pi i \tau}$, and $\mathbf{L}_0$ is the generator of the algebra satisfying $\mathbf{L}_0|\tilde{\Delta}\rangle =\tilde{\Delta}|\tilde{\Delta}\rangle$. Here, $\mathcal{F}(\Delta_1, \tilde{\Delta}, q)_A$ is the one-point holomorphic torus CB (for more details, see, e.g., \cite{He:2012bi, Hadasz:2009db, DiFrancesco:1997nk}). In the case of the Virasoro algebra and generic central charge $c$, an exact expression for $\mathcal{F}(\Delta_1, \tilde{\Delta}, q)_A$ is unknown. When one restricts the analysis to the large central charge limit, exact expressions are known for the $\mathfrak{sl}_2$ \textit{global} one-point\footnote{For global multi-point blocks, some expressions are known in specific channels, see \cite{Alkalaev:2016fok, Alkalaev:2022kal, Pavlov:2023asi}.} torus CBs. \textit{Global conformal blocks} are defined as the contribution of CBs coming from the $\mathfrak{sl}_N$ subalgebra of the $\mathcal{W}_N$ algebra. For other $\mathcal{W}_N$ ($N\ge3$) conformal field theories, no exact expressions are currently known. In \cite{Belavin:2023orw}, a perturbative expression was presented for $\mathcal{W}_3$ CFT in the large central charge limit. In this work, we derive the exact expression for the $\mathfrak{sl}_3$ global one-point CBs, which is the main result of this paper.

We consider the light operators relevant for the global CBs, whose conformal dimensions scale as $\Delta \sim \mathit{o}(1)$ as $c \to \infty$. This behavior allows to restrict the set of generators of $\mathcal{W}_2$ and $\mathcal{W}_3$ to those of $\mathfrak{sl}_2$ and $\mathfrak{sl}_3$ algebras, respectively, and leads to a significant simplification of the CBs.

$\mathcal{W}_3$ primary fields $\phi_{j}(z, \bar{z})$ are labeled by a vector $j$ belonging to the root space of $\mathfrak{sl}_3$,
\begin{equation} \label{vofsl3}
j=r w_1+s w_2,
\end{equation}
where $w_1$, $w_2$ are $\mathfrak{sl}_3$ fundamental weights. Unlike $\mathcal{W}_2$, in $\mathcal{W}_3$ CFT, the CBs are not fully determined by the symmetry algebra due to the presence of multiplicities in the OPE of primary fields. In the language of $\mathfrak{sl}_3$ representation, this translates into multiplicities in the tensor product of $\mathfrak{sl}_3$ representations. Therefore, below, we will restrict our discussion to a class of CBs for which the problem of multiplicities is absent~\cite{Fateev:2007ab}, with a number of external fields fulfilling the following condition
\begin{equation} \label{alpha4}
    \bW_{-1}\phi_{j} (0,0)\ket{0}= \frac{3q_j}{2\Delta_j}\bL_{-1}\phi_{j}(0,0)\ket{0}\;.
\end{equation}
Here $\Delta_j, q_j$ are the conformal dimension and $\mathcal{W}_3$ \textit{charge} of $\phi_{j}$, respectively.

The paper is organized as follows. In section \bref{firstsection}, we briefly review some basic concepts of Virasoro CFT. In section \bref{basiccon1}, we explain the fundamental elements of the shadow formalism in the $\mathfrak{sl}_2$ case. 
We introduce a shadow operator that allows us to express CBs in terms of conformal partial waves. 
We discuss this construction for the sphere topology in section \bref{sphereCBPW} and for the torus topology in section \bref{1ptpw1}. In section \bref{sl3blocksection}, we recall the basic facts about $\mathcal{W}_3$ CFT, define the $\mathfrak{sl}_3$ global one-point torus block and provide its perturbative expression. Section \bref{geralsl3shf} is devoted to generalizing the shadow formalism to the $\mathfrak{sl}_3$ case. In section \bref{subsecsl3invf}, we introduce preliminary concepts related to the $\mathfrak{sl}_3$ invariant functions theory. In section \bref{sl3shadowformconst}, we generalize the shadow formalism to the $\mathfrak{sl}_3$ case. In section \bref{sl34ptsphereCB}, we apply the constructed formalism to the computation of the $\mathfrak{sl}_3$ global four-point sphere CB, and in section \bref{sl3globalonepointtorus}, we apply the formalism to compute the $\mathfrak{sl}_3$ global one-point torus CB. In section \bref{conclusions}, we present our conclusions. Appendices \bref{app.CI} and \bref{thesl3integral} are included to explain some technical details of certain integral computations. In appendix \bref{apsl3fields}, we present a short review of $\mf{sl}_3$ fields, which are introduced in section \bref{sl3shadowformconst}.

\section{$\mf{sl}_2$ global conformal blocks via shadow formalism}  \label{firstsection}
\flushbottom
The symmetry of the Virasoro CFT is generated by the energy-momentum tensor $\mathbf{T}(z)$\footnote{Similar discussion holds for antiholomorhic $\bar{T}(\bar{z})$} (a spin-$2$ current) whose Laurent series expansion reads
\begin{equation}  \label{sl2gbs1}
\mathbf{T}(z)= \sum_{n=-\infty}^{\infty} \frac{\bL_n}{z^{n+2}}.
\end{equation}
The modes $\bL_n$ satisfy the Virasoro algebra 
\begin{equation} \label{sl2gbs2}
 \begin{split}
\left[\bL_n, \bL_m\right]		& =(n-m) \bL_{n+m}+\dps \frac{c}{12}(n^3-n) \delta_{n+m,0} \; .
\end{split}
\end{equation}
We denote by $\varphi_{\Delta, \bar{\Delta}} (z, \bar{z})$ the Virasoro primary fields with conformal dimensions $\Delta, \bar{\Delta}$. 
For simplicity, in what follows, we assume that holomorphic and antiholomorphic conformal dimensions are equal
\begin{equation} \label{sl2gbs3}
    \Delta=\bar{\Delta},
\end{equation}
and denote the primary fields
\begin{equation} \label{sl2gbs4}
    \varphi_{\Delta} (z, \bar{z}):=  \varphi_{\Delta, \bar{\Delta}} (z, \bar{z}).
\end{equation}
One can show that in the limit $c \to \infty$,  in order to have finite inner product  of states $\bra{\Delta,n}   n,\Delta\rangle$, where $ \ket{n,\Delta}=\bL_{-n}\varphi_{\Delta}(0,0)\ket{0}$, one needs to restrict the generators $\bL_{n}$ to the set $\bL_0, \bL_1, \bL_{-1}$, which form the $\mf{sl}_2$ subalgebra
\begin{equation} \label{sl2gbs5}
    \left[\bL_n, \bL_m\right]	 =(n-m) \bL_{n+m}.
\end{equation}
The generators $\bL_n$ satisfy the following commutation relations with the primary fields
\begin{equation} \label{sl2gbs6}
    [\bL_n, \varphi_{\Delta}(z, \bar{z})]= \mathcal{L}_n \varphi_{\Delta} (z,\bar{z}),  \quad \mathcal{L}_n= z^n(z \partial_z+\Delta(n+1)).
\end{equation}
The differential operators $\mc{L}_0, \mc{L}_1, \mc{L}_{-1}$ are the generators of $\mf{sl}_2$ transformations on primary fields
\begin{equation} \label{sl2tranfor}
\mc{L}_0= z \partial_z+ \Delta,\quad \mc{L}_1= z^2 \partial_z+ 2 z \Delta, \quad \mc{L}_{-1}=  \partial_z.
\end{equation}
To compute CBs on sphere and torus, one uses the OPE decomposition of the product of primary fields and the commutation relations (\ref{sl2gbs6}) to compute matrix elements of the type 
\begin{equation} \label{sl2gbs7}
      \bra{\Delta_1, N}\varphi_{\Delta_2}(z, \bar{z})  \ket{N, \Delta_3},
\end{equation}
which arise in the decompositions of the CBs. Here $\ket{N, \Delta_i}$  stands for the descendent states of $\ket{\Delta_i}$. On the sphere at large $c$ limit, the CBs get contributions only from $\mf{sl}_2$ generators; thus, the CBs reduce to the global CBs. While on the torus, besides the global blocks, one also has the so-called \textit{light} CBs, which contain contributions from the full Virasoro generators. In this work, we will concentrate only on the global blocks. To compute them, we use the shadow formalism.

\subsection{Basics of $\mathfrak{sl}_2$ Shadows formalism} \label{basiccon1}
\hfill\break
\textbf{Basic concepts} 
\hfill\break
Let us introduce the shadow operator of the primary field $\varphi_{\Delta} (z, \bar{z})$. Taking into account (\ref{sl2gbs3}), the shadow operator is defined as
\begin{equation} \label{shadop1}
 \tilde  { \varphi}_{\Delta^{*}} (z, \bar{z})= \frac{1}{\mathcal{N}_{\Delta} } \int_{\mathbb{R}^2} d^2 w \frac{\varphi_{\Delta}(w, \bar{w})}{|z-w|^{4 (1-\Delta) } },
\end{equation}
where $d^2 w=dw d\bar{w}$ represents integration over the complex plane, and $\Delta^{*}$ is the holomorphic and antiholomorphic conformal dimension of the shadow operator (which is also a primary field)
\begin{equation} \label{sl2gbs8}
    \Delta^{*}=1-\Delta,
\end{equation}
and $\mathcal{N}_{\Delta}$ is a normalization coefficient
\begin{equation} \label{thecoefN}
    \mathcal{N}_{\Delta} = \pi^2    \frac{\Gamma(2\Delta-1) \Gamma(1-2 \Delta) }{ \Gamma(2-2 \Delta )  \Gamma(2  \Delta )   }.
\end{equation}
Below, we will show that the shadow operator has the property
\begin{equation} \label{deltafun}
\langle  \tilde{\varphi}_{\Delta^{*}}(\ocom{z})  \varphi_{\Delta}(\ocom{w}) \rangle = \delta^2 (z-w),
\end{equation}
where the two-dimensional delta function $\delta^2(z-w)$ is defined according to $\int_{\mathbb{R}^2} f (z, \bar{z})\delta^2 (z-w) d^2z =f(w, \bar{w})$. We define the ``projector'' operator
\begin{equation} \label{sl2gbs9}
    \Pi_{\Delta} = \int_{\mathbb{R}^2}    d^2z  \varphi_{\Delta}(\ocom{z}) \ket{0}\bra{0}  \tilde{\varphi}_{\Delta^{*}}(\ocom{z}).
\end{equation}
It satisfies the property
\begin{equation} \label{sl2gbs10}
    \Pi _{\Delta_m} \Pi_{\Delta_n}= \delta_{m n}\Pi_{\Delta_m   }.
\end{equation}
For the operator 
\begin{equation}  \label{sl2gbs11}
\mathcal{P}= \sum_{ \Delta \in \mathcal{D} } \Pi_{\Delta},
\end{equation}
where $\mathcal{D}$ is the domain of admissible conformal dimensions for the considered CFT, one can show that
\begin{equation} \label{iprosl2}
    \mathcal{P}\ket{h} = \ket{h}, \quad \text {where $ \ket{h}= \varphi_h (0,0) \ket{0} $}.
\end{equation}
This property follows directly from the definition (\ref{sl2gbs9}, \ref{sl2gbs11}). By writing explicitly (\ref{sl2gbs11}), we have
\begin{equation} \label{ipros13}
    \mathcal{P} \ket{h} = \sum_{\Delta \in \mathcal{D}} \int_{\mathbb{R}^2}    d^2z  \varphi_{\Delta}(z, \bar{z}) \ket{0}\bra{0}  \tilde{\varphi}_{\Delta^{*}}(z, \bar{z}) \varphi_{h}(0,0) \ket{0}.
\end{equation}
Because of  (\ref{deltafun}), we have
\begin{equation} \label{2ptcases}
  \bra{0}\tilde{\varphi}_{\Delta^*}(z, \bar{z})  \varphi_h(0,0) \ket{0} = \begin{cases}\delta_{h, \Delta} \delta^2(z), \\ 0, \quad \text{otherwise}.\end{cases}
\end{equation}
Hence
\begin{equation} \label{actionofI}
     \mathcal{P} \ket{h} = \sum_{\Delta \in \mathcal{D}} \int_{\mathbb{R}^2}    d^2z  \varphi_{\Delta}(z, \bar{z}) \ket{0} (  \delta_{h, \Delta} \delta^2(z))= \varphi_h(0,0)\ket{0} = \ket{h}.
\end{equation}
We notice that since $\tilde{\varphi}_{\Delta^*}(z, \bar{z})$ is a primary field, one could include an extra contribution to the two-point correlation function
\begin{equation}  \label{2ptcases2}
   \bra{0}\tilde{\varphi}_{\Delta^*}(z, \bar{z})  \varphi_h(0,0) \ket{0} = \begin{cases}\delta_{h, \Delta} \delta^2 (z), \\   \delta_{h, 1-\Delta}\frac{1}{z^{2(1-\Delta)} \bar{z}^{2(1-\Delta)}  }, \\ 0, \quad \text{otherwise} .  \end{cases}
\end{equation}
However, this modification does not substantially change (\ref{actionofI}). 
Indeed, assuming (\ref{2ptcases2}), we have
\begin{equation} \label{actionofI2}
\begin{split}
     &\mathcal{P} \ket{h} = \sum_{\Delta \in \mathcal{D}} \int_{\mathbb{R}^2}    d^2z \varphi_{\Delta}(z, \bar{z}) \ket{0}\bra{0}  \tilde{\varphi}_{\Delta^*}(z, \bar{z}) \varphi_{h}(0,0) \ket{0}=\\& \sum_{\Delta \in \mathcal{D}} \int_{\mathbb{R}^2}    d^2z \varphi_{\Delta}(z, \bar{z}) \ket{0} (  \delta_{h, \Delta} \delta^2(z)+\delta_{1-\Delta,h} \frac{1}{z^{2(1-\Delta)}  \bar{z}^{2(1-\Delta)} })= \\&   \ket{h} + \int d^2z   \frac{\varphi_{1-h}(z, \bar{z})}{z^{2h}  \bar{z}^{2h}}\ket{0}= \ket{h}+\mathcal{N}_{1-h}\tilde{\varphi}_{h}(0,0)\ket{0}= (1+\mathcal{N}_{1-h}) \ket{h}\propto \ket{h},
 \end{split}
\end{equation}
where in the third line we used that $\tilde{\varphi}_h$ can be expressed in terms of its shadow field $\varphi_{h^*}$. The result (\ref{actionofI2}) is essentially the same as (\ref{actionofI}).

\hfill\break
\textbf{Delta function}
\hfill\break
The relation (\ref{deltafun}) can be established as follows. By inserting the shadow operator (\ref{shadop1}) into the lhs of (\ref{deltafun}), and writing it for $z_1$ and $z_2$ coordinates, we have

\begin{equation} \label{ipros14}
    \langle\tilde{\varphi}_{\Delta^*} (z_2, \bar{z}_2) \varphi_{\Delta} (z_1, \bar{z}_1)  \rangle = \frac{1}{\mc{N}_{\Delta}} \int d^2 z  \frac{  \langle \varphi_{\Delta}(z, \bar{z})  \varphi_{\Delta}(z_1, \bar{z}_1) \rangle}{|z-z_2|^{4(1-\Delta)}} = \delta^2 (z_1-z_2).
\end{equation}
Since $\langle \varphi_{\Delta}(z, \bar{z}) \varphi_{\Delta}(z_1, \bar{z}_1) \rangle = \frac{1}{|z-z_1|^{4 \Delta}}$, then (\ref{ipros14}) becomes
\begin{equation} \label{ipros15}
    \frac{1}{\mc{N}_{\Delta}} \int d^2 z  \frac{  1}{|z-z_2|^{4(1-\Delta)} |z-z_1|^{4 \Delta}} = \delta^2 (z_1-z_2).
\end{equation}
By using the parametrization $h_1=2 \Delta$, $h_2=2-h_1= 2(1-\Delta)$, eq.~(\ref{ipros15}) can be written more generally as
\begin{equation} \label{ipros16}
    \frac{1}{\mc{N}_{\Delta}} \int d^2 z  \frac{1}{(z-z_1)^{h_1}  (z-z_2)^{h_2}} \frac{1}{(\bar{z}-\bar{z}_1)^{h_1}  (\bar{z}-\bar{z}_2)^{h_2}}= \delta^2(z_1-z_2),
\end{equation}
where $\mathcal{N}_{\Delta}$ is expressed in terms of $h_1, h_2$ as
\begin{equation} \label{ipros17}
   \mc{N}_{\Delta}=  \frac{\pi^2 \Gamma({1-h_1}) \Gamma({1-h_2}) }{  \Gamma({h_1})  \Gamma({h_2}) } .
\end{equation}
Let us prove (\ref{ipros15}) and show that $\mathcal{N}_{\Delta}$ is given by (\ref{thecoefN}) or (\ref{ipros17}). For this purpose, we write (\ref{ipros15}) in Cartesian coordinates, and use the above parametrization $h_1, h_2$, then (\ref{ipros15}) becomes
\begin{equation} \label{pofdf1}
     \frac{1}{\mc{N}_{\Delta} }  \int d^2x \frac{1}{(x-x_1)^{2 h_1} (x-x_2)^{2 h_2} }= \delta^2 (x_1-x_2).
\end{equation}
By using the relation
\begin{equation}
    \frac{\Gamma(\Delta)}{x^{2 \Delta}} = \int_{0}^{\infty} dt_1  t_1^{\Delta-1} e^{-t_1 x^2},
\end{equation}
we rewrite (\ref{pofdf1}) as follows   
\begin{equation}
\begin{split}
& \frac{1}{\mc{N}_{\Delta} }  \int d^2x \frac{1}{(x-x_1)^{2 h_1} (x-x_2)^{2 h_2} } = \\& \frac{1}{ \mc{N}_{\Delta} \Gamma({h_1})  \Gamma({h_2})}\int d^2x \int_{0}^{\infty} dt_1 dt_2  t_1^{h_1-1}  t_2^{h_2-1}e^{-t_1 (x-x_1)^2-t_2 (x-x_2)^2}.
\end{split}
\end{equation}
After some transformations and integration over $x$, one can convert this integral to
    \begin{equation}
\begin{split} \label{pofdel2}
& \frac{1}{\mc{N}_{\Delta} }  \int d^2x \frac{1}{(x-x_1)^{2 h_1} (x-x_2)^{2 h_2} } = \\& \frac{\pi}{ \mc{N}_{\Delta} \Gamma({h_1})  \Gamma({h_2})} \int_{0}^{\infty} \frac{dt_1 dt_2}{t_1+t_2}  t_1^{h_1-1}  t_2^{h_2-1}e^{-\frac{t_1t_2(x_1-x_2)^2}{t_1+t_2}}.
\end{split}
\end{equation}
For the exponent, we use the Fourier transform
\begin{equation}
e^{-\alpha x^2} = \int d^2k    \left( e^{-\frac{k^2}{4 \alpha}} \frac{\pi}{\alpha}   \right) e^{-i k\cdot x}.
\end{equation}
By multiplying the rhs of (\ref{pofdf1}) by $e^{i k \cdot (x_1-x_2)}$ and performing the two-dimensional integral over the variable $(x_1-x_2)$, we obtain 1. Therefore, by multiplying both sides of (\ref{pofdel2}) and integrating over $(x_1-x_2)$, we must obtain the same result. Applying this reasoning, we obtain from (\ref{pofdel2})
\begin{equation}
    \frac{\pi^2}{ \mc{N}_{\Delta} \Gamma({h_1})  \Gamma({h_2}) } \int_0^{\infty}  dt_1 t_1^{h_1-1} e^{-\frac{k^2}{ 4 t_1}}  \int_0^{\infty}  dt_2 t_2^{h_2-1} e^{-\frac{k^2}{ 4 t_2}}=1.
\end{equation}
This implies that
\begin{equation}
    \frac{\pi^2 \Gamma({1-h_1}) \Gamma({1-h_2}) }{ \mc{N}_{\Delta} \Gamma({h_1})  \Gamma({h_2}) }  =1\;,
\end{equation}
where we used that $h_1+h_2=2$. Hence 
\begin{equation}
   \mc{N}_{\Delta}=  \frac{\pi^2 \Gamma({1-h_1}) \Gamma({1-h_2}) }{  \Gamma({h_1})  \Gamma({h_2}) } ,
\end{equation}
in accordance with (\ref{thecoefN}) and (\ref{ipros17}).

\subsection{Sphere: Conformal blocks and partial waves} \label{sphereCBPW}
On the sphere, the four-point correlation function can be decomposed into four-point sphere CBs $\mathcal{F}^{s}( \mathbf{\Delta}_4, \tilde{\Delta}, \mathbf{z}_4)$ as follows\footnote{Here, we omit the standard prefactor, which is fixed by the global Ward identities.}
\begin{equation} \label{spblock1}
  G^{s}_{\mathbf{\Delta}_4}( \mathbf{z}_4, \bar{ \mathbf{z}}_4)=\langle \prod_{i=1}^4 \varphi_{\Delta_i} (z_i, \bar{z}_i)  \rangle  = \sum_{\tilde{\Delta} \in \mc{D}} C_{\Delta_1 \Delta_2 \tilde{\Delta}}  C_{\tilde{\Delta} \Delta_3 \Delta_4}|\mc{F}^s( \mathbf{\Delta}_4, \tilde{\Delta}, \mathbf{ z}_4)|^2,
\end{equation}
where $\mathbf{\Delta}_n=\Delta_1,\Delta_2, ..., \Delta_n$, $\mathbf{z}_n= z_1, z_2,...,z_n $. In the large $c$ limit $\mc{F}^s( \mathbf{\Delta}_4, \tilde{\Delta}, \mathbf{ z}_4)$ becomes the $\mf{sl}_2$ global four-point sphere CB $\mc{F}^s( \mathbf{\Delta}_4, \tilde{\Delta}, \mathbf{ z}_4)_{\mf{sl}_2}$ given by
\begin{equation} \label{spblock2}
    \mc{F}^s( \mathbf{\Delta}_4, \tilde{\Delta}, \mathbf{ z}_4)_{\mf{sl}_2}  = x^{\tilde{\Delta}- \Delta_1-\Delta_2} {}_2F_1 ( \tilde{\Delta} -\Delta_{12}, \tild+\Delta_{34}, 2 \tild, x),
\end{equation}
where $x$ is the cross-ratio $x= \frac{z_{12} z_{34}}{ z_{13} z_{24}} $ and ${}_2F_1(a,b,c,x)$ denotes the hypergeometric function. In the large $c$ limit, one can use the shadow formalism to decompose (\ref{spblock1}) into partial waves. This is done by inserting the resolution of identity operator $\mathcal{P}$ in (\ref{spblock1}) as follows
\begin{equation} \label{spblock3}
\begin{split}
     & G^{s}_{\mathbf{\Delta}_4}( \mathbf{z}_4, \bar{ \mathbf{z}}_4)=\langle \varphi_{\Delta_1} (z_1, \bar{z}_1)  \varphi_{\Delta_2} (z_2, \bar{z}_2) \mc{P} \varphi_{\Delta_3} (z_3, \bar{z}_3) \varphi_{\Delta_4} (z_4, \bar{z}_4) \rangle   =\\& \sum_{\tild \in \mc{D}}
     \int_{\mathbb{R}^2}    d^2z      \langle \varphi_{\Delta_1} (z_1, \bar{z}_1)  \varphi_{\Delta_2} (z_2, \bar{z}_2)  \varphi_{\tilde{\Delta}}(\ocom{z}) \ket{0}\bra{0}  \tilde{\varphi}_{\tilde{\Delta}^{*}}(\ocom{z}) \varphi_{\Delta_3} (z_3, \bar{z}_3) \varphi_{\Delta_4} (z_4, \bar{z}_4) \rangle  .
\end{split}
\end{equation}
The first factor of the integrand is just the three-point function, generally given by
\begin{equation} \label{3ptfunction2}
    \langle   \varphi_{\Delta_1} (z_1, \bar{z}_1)  \varphi_{\Delta_2} (z_2, \bar{z}_2)  \varphi_{\Delta_3} (z_3, \bar{z}_3)    \rangle  =  C_{\Delta_1\Delta_2 \Delta_3 }   |V_{\Delta_1, \Delta_2, \Delta_3}(z_1, z_2, z_3)|^2,
\end{equation}
where $V_{\Delta_1, \Delta_2, \Delta_3}(z_1, z_2, z_3)$ is the coordinate dependence 
\begin{equation} \label{3ptcoord}
    V_{\Delta_1, \Delta_2, \Delta_3}(z_1, z_2, z_3)= \frac{1}{z_{12}^{\Delta_1+\Delta_2-\Delta_3} z_{13}^{\Delta_1+\Delta_3-\Delta_2} z_{23}^{\Delta_2+\Delta_3-\Delta_1}},
\end{equation}
where $z_{ij}=z_i-z_j$. In the next section, we will prove that the second factor in the integrand  $\bra{0}  \tilde{\varphi}_{\tilde{\Delta}^{*}}(\ocom{z})\varphi_{\Delta_3} (z_3, \bar{z}_3) \varphi_{\Delta_4} (z_4, \bar{z}_4) \rangle$ is also proportional to the three-point function $|V_{1-\tilde{\Delta}, \Delta_3, \Delta_4}(z, z_3, z_4)|^2$. Thus, (\ref{spblock3}) can be written as
\begin{equation} \label{spblock4}
    G^{s}_{\mathbf{\Delta}_4}( \mathbf{z}_4, \bar{\mathbf{z}}_4) =\sum_{\tild \in \mc{D}} B(\mathbf{\Delta}_4, \tild)    \int_{\mathbb{R}^2}    d^2z     |V_{ \Delta_1, \Delta_2, \tilde{\Delta}}(z_1, z_2, z)|^2  |V_{1-\tilde{\Delta}, \Delta_3, \Delta_4}(z, z_3, z_4)|^2,
\end{equation}
where $B(\mathbf{\Delta}_4, \tild) $ is proportional to $C_{\Delta_1 \Delta_2 \tilde{\Delta}}  C_{\tilde{\Delta} \Delta_3 \Delta_4}$. The object on the rhs of (\ref{spblock4})

\begin{equation} \label{spblock5}
    \Psi_{\tild}^{\mathbf{\Delta}_4} (\mathbf{z}_4, \bar{ \mathbf{z}}_4)= \int_{\mathbb{R}^2}  d^2z     |V_{ \Delta_1, \Delta_2, \tilde{\Delta}}(z_1, z_2, z)|^2  |V_{1-\tilde{\Delta}, \Delta_3, \Delta_4}(z, z_3, z_4)|^2
\end{equation}
is termed the \textit{four-point sphere conformal partial wave}. Equations (\ref{spblock1}, \ref{spblock2}, \ref{spblock4}, \ref{spblock5}) show the direct relation between $\mathfrak{sl}_2$ global four-point sphere CBs and the four-point sphere partial waves. Indeed, if one computes the integral (\ref{spblock5}) (for details of the computation of this integral, see appendix A of \cite{Dolan:2011dv}), one finds that the four-point partial wave is given by a linear combination of two terms $|\mathcal{F}^s( \mathbf{\Delta}_4, \tilde{\Delta}, \mathbf{z}_4)_{\mathfrak{sl}_2}|^2$, $|\mathcal{F}^s( \mathbf{\Delta}_4, 1-\tilde{\Delta}, \mathbf{z}_4)_{\mathfrak{sl}_2}|^2$.

Similar relationships are observed between partial waves and higher-point CBs in the comb channels. In these cases, one has $(n+2)$-point correlation functions 
\begin{equation} \label{n2ptcorrefunc}
    \langle  \prod_{i=1}^{n+2} \varphi_{\Delta_i} (z_i, \bar{z}_i)\rangle\;,
\end{equation}
which are decomposed in the comb channel into $(n+2)$-point sphere CBs. Using a decomposition analogous to the one previously discussed, the correlation functions (\ref{n2ptcorrefunc}) can also be expressed in terms of $(n+2)$-point partial waves, defined as
\begin{equation}
\begin{split}
   \Psi_{\mathbf{\tild}_{n-1}}^{\mathbf{\Delta}_{n+2}} (\mathbf{z}_{n+2}, \bar{\mathbf{z}}_{n+2})= \int_{\mathbb{R}^{2(n-1)}}  \prod_{i=1}^{n-1} & d^2w_i     |V_{ \Delta_1, \Delta_2, \tilde{\Delta}_1}(z_1, z_2, w_1)|^2  |V_{\tilde{\Delta}_1^*, \Delta_3,  {\tilde{\Delta}_2}}(w_1, z_3, w_2)|^2...\times \\& ...\times |V_{\tilde{\Delta}_{n-1}^*, \Delta_{n+1}, \Delta_{n+2}} (w_{n-1}, z_{n+1}, z_{n+2})|^2.
   \end{split}
\end{equation}
This implies a direct relation between the CBs and the partial waves (for a more detailed discussion, see \cite{Rosenhaus:2018zqn}).

\subsection{Torus} \label{1ptpw1}
\subsubsection{One-point torus conformal blocks via shadow formalism} \label{1ptpw1.2}
The one-point torus correlation function is given by (\ref{main1ptblock}), where in the case of the Virasoro CFT, CBs $\mathcal{F}(\Delta_1, \tilde{\Delta},q)_A$ (where $A$ denotes the Virasoro algebra) receive contributions from the full Virasoro algebra. We are interested in the contribution from the $\mathfrak{sl}_2$ subalgebra. This contribution is given by the $\mathfrak{sl}_2$ global one-point torus block $\mathcal{F}( \Delta_1, \tilde{\Delta},q)_{\mathfrak{sl}_2}$, defined as follows:

\begin{equation} \label{1ptexpansion}
    \mathcal{F}( \Delta_1, \tilde{\Delta},q)_{\mf{sl}_2} =\frac{1}{\tilde{C}_{\tild \Delta_1  \tild}}\sum_{  m=0  }   \left( B_{\tilde{\Delta}}^{-1}   \right)^{ mm}   \bra{\tilde{\Delta},m}  \varphi_{\Delta_1}(z_1, \bar{z}_1) q^{\bL_0} \ket{m, \tilde{\Delta}},
\end{equation}
where
\begin{equation} \label{1ptcor4}
    \ket{m,\tild} = (\bL_{-1})^m \varphi_{\tild}(0,0) \ket{0} , \quad B_{\tild}^{ mm} = \bra{\tild, m} m, \tild \rangle, 
\end{equation}
and $ \left( B_{\tilde{\Delta}}^{-1}   \right)^{ mm} $ is the inverse of $B_{\tild}^{ mm}$, $\tilde{C}_{\tild \Delta_1 \tild}=\bra{\tild} \varphi_{\Delta_1} (z_1, \bar{z}_1) \ket{\tild}=z_1^{-\Delta_1} \bar{z}_1^{-\Delta_1} C_{\tild \Delta_1 \tild}$.\footnote{We notice that the dependence on $z_1$ of this matrix element disappears after transforming to cylinder coordinates.} For simplicity of writing, we assume that the normalization factor $\tilde{C}_{\tild \Delta_1 \tild}=1$. We aim to show that, similar to the spherical case, the block (\ref{1ptexpansion}) can be expressed in terms of one-point torus partial waves defined in some integral representation. Our discussion here follows \cite{Alkalaev:2023evp}, though the treatment is slightly different. The basic object used to decompose the rhs of (\ref{1ptexpansion}) into partial waves is the resolution of identity operator $\mathcal{P}$ (\ref{sl2gbs11}) constructed in the previous section. Let us insert $\mathcal{P}$ into (\ref{1ptexpansion})

\begin{equation} \label{1ptcor5}
\begin{split}
    &\mathcal{F}( \Delta_1,\tilde{\Delta}, q)_{\mf{sl}_2} =\sum_{  m=0  }   \left( B_{\tilde{\Delta}}^{-1}   \right)^{ mm}   \bra{\tilde{\Delta},m}  \mathcal{P} \varphi_{\Delta_1}(z_1, \bar{z}_1)  q^{\bL_0} \ket{m, \tilde{\Delta}}=   \\&
\sum_{  m=0  }   \left( B_{\tilde{\Delta}}^{-1}   \right)^{ mm}   \bra{\tilde{\Delta},m}  \left(   \sum_{h}  \int d^2z   \varphi_{h}(z, \bar{z}) \ket{0}  \bra{0}\tilde{\varphi}_{h^*} (z, \bar{z})  \right) \varphi_{\Delta_1}(z_1, \bar{z}_1)  q^{\bL_0} \ket{m, \tilde{\Delta}}.
\end{split}
\end{equation}
In the factor $\bra{\tild,m} \varphi_h(z, \bar{z}) \ket{0}$ by utilizing the formulas 
\begin{equation} \label{expfor}
\begin{split}
& \varphi_{\tilde{\Delta}}(z, \bar{z})  = e^{z \bL_{-1} + \bar{z}\bar{\bL}_{-1}} \varphi_{\tilde{\Delta}}(0,0)  e^{-z \bL_{-1} -\bar{z} \bar{\bL}_{-1}},\\
&\varphi_{\tilde{\Delta}}(z , \bar{z}) \ket{0} = e^{z \bL_{-1} + \bar{z}\bar{\bL}_{-1}} \varphi_{\tilde{\Delta}}(0,0)  e^{-z \bL_{-1} -\bar{z} \bar{\bL}_{-1}} \ket{0} =  \sum_{n, \bar{n}=0} \frac{ z^n \bar{z}^{\bar{n}} }{n! \bar{n}!}  \bL_{-1}^n \bar{\bL}_{-1}^{\bar{n}}  \varphi_{\tilde{\Delta}}(0,0)\ket{0},
    \end{split}
\end{equation}
one obtains 
\begin{equation} \label{matrixE0}
   \bra{\tild,m} \varphi_h(z, \bar{z}) \ket{0} = \delta_{h, \tild}    \frac{z^m}{m!} B_{\tilde{\Delta}}^{mm} . 
\end{equation}
Taking into account (\ref{matrixE0}) and writing $\tilde{\varphi}_{h^*}$ according to  (\ref{shadop1}), (\ref{1ptcor5}) becomes
\begin{equation} \label{1ptcor6}
\begin{split}
\mathcal{F}( \Delta_1,\tilde{\Delta} , q)_{\mf{sl}_2}=    \sum_{  m=0  } q^{\tilde{\Delta}+m}  & \int \frac{d^2z d^2w}{\mathcal{N}_{\tild} |z-w|^{4(1-\tild)}}   \frac{z^{m}}{m!} \bra{0} \varphi_{\tild}(w,\bar{w})    \varphi_{\Delta_1}(z_1, \bar{z}_1)    \ket{ m, \tild} .
\end{split}
\end{equation}
Using
\begin{equation} \label{expansiontrans}
    \varphi_{\tild}(zq, 0)\ket{0} = \sum_{m=0} \frac{z^m q^m}{m!}   \ket{ m, \tild},
\end{equation}
we obtain
\begin{equation} \label{blocketo3}
     \mathcal{F}(\Delta_1, \tilde{\Delta},q)_{\mf{sl}_2} = q^{\tilde{\Delta}}  \int d^2z  d^2w   \frac{\bra{0}  \varphi_{\tilde{\Delta}}( w,\bar{w})   \varphi_{\Delta_1}(z_1, \bar{z}_1)\varphi_{\tild} (z q , 0)\ket{0} } { \mathcal{N}_{\tild } |z-w|^{4 (1-\tilde{\Delta)} }}.
\end{equation}
To write explicitly the integrand, we use (\ref{3ptfunction2}) (for the moment, we write only the coordinate dependence). Thus, (\ref{blocketo3}) becomes
\begin{equation}
\begin{split}
& \mathcal{F}( \Delta_1,\tilde{\Delta}, q)_{\mf{sl}_2} = \frac{q^{\tild}}{\mathcal{N}_  {\tilde{\Delta}}}
  \int d^2z d^2w     \frac{1}{  (z-w)^{2(1-\tild)} (\bar{z}-\bar{w})^{2 (1-\tild)} (w-z_1)^{\Delta_1} (\bar{w}-\bar{z}_1)^{\Delta_1} } \times \\ \times &\frac{1}{  (w-zq)^{2 \tild -\Delta_1}  \bar{w}^{2 \tild -\Delta_1} (z_1-z q)^{\Delta_1}  \bar{z}_1 ^{\Delta_1} }  .
\end{split}
\end{equation}
Integrating over $w, \bar{w}$ and using the formula (\ref{cint3}), we obtain 

\begin{equation} \label{integral1}
\begin{split}
\mathcal{F}(\Delta_1, \tilde{\Delta},q)_{\mf{sl}_2} =C_2 q^{\tild}  \int d^2z V_{1-\tilde{\Delta},  \Delta_1, \tilde{\Delta}}(z,z_1,zq)  V_{1-\tilde{\Delta},  \Delta_1, \tilde{\Delta}}(\bar{z},\bar{z}_1,0) ,
\end{split}
\end{equation}
where       
\begin{equation}
    C_2= \frac{\pi K_{123} (2-2\tild, \Delta_1, 2\tild-\Delta_1)}{\mathcal{N}_ {\tilde{\Delta} } } ,
\end{equation}
and $K_{123} (2-2\tild, \Delta_1, 2\tild-\Delta_1)$ is according to (\ref{K123}). Equation (\ref{integral1}) is the main formula we wanted to prove in this section. The rhs of (\ref{integral1}), up to the factor $C_2$, is the holomorphic contribution in the variable $q$ to the one-point torus partial wave. We will discuss it in more detail below. In order to verify (\ref{integral1}), we notice first  that the lhs is given by
\begin{equation} \label{expl1pt1}
\begin{split}
  \mathcal{F}( \Delta_1,\tilde{\Delta}, q)_{\mf{sl}_2} = & \frac{q^{\tilde{\Delta}}}{(1-q)^{1-\Delta_1}} {}_2F_1 (  \Delta_1, \Delta_1+2\tild-1, 2 \tild,q) =\\&
   q^{\tild} \left( 1+  q( 1+  \frac{\Delta_1(\Delta_1-1)}{2\tild}   ) +... \right)  .
    \end{split}
\end{equation}
Now, the rhs of (\ref{integral1})
\begin{equation} \label{integral5}
\begin{split}
    &C_2 q^{\tilde{\Delta}}  \int d^2 z V_{1-\tilde{\Delta},  \Delta_1, \tilde{\Delta}}(z,z_1,zq)  V_{1-\tilde{\Delta},  \Delta_1, \tilde{\Delta}}(\bar{z},\bar{z}_1,0)  =  \\ & C_2 q^{\tild} \int   \frac{d^2 z}{  (z-z_1)^{\Delta_1-2 \tild+1} (z-zq )^{1-\Delta_1} (z_1-z q)^{2 \tild +\Delta_1-1} (\bar{z}-\bar{z}_1)^{-2\tild +\Delta_1+1} (\bar{z})^{1-\Delta_1} } \times \\&  \times  \frac{1}{   (\bar{z}_1 )^{2 \tild +\Delta_1-1} }
    = \\ & C_2\frac{q^{\tild} z_1^{-\Delta_1} \bar{z}_1^{-\Delta_1}}{(1-q)^{1-\Delta_1}} \int  \frac{d^2 z} {(1-z)^{\Delta_1-2\tild+1}   z^{1-\Delta_1}  (1-z q)^{2 \tild+\Delta_1-1}  (1-\bar{z})^{\Delta_1-2\tild+1 }  \bar{z}^{1-\Delta_1} },
\end{split}
\end{equation}
where in the last line, we did a change of variables $z \rightarrow z_1 z, \bar{z} \to \bar{z}_1 \bar{z}$. Expanding in $q$ the integrand we have
\begin{equation}
\begin{split} \label{integral6}
    & C_2 q^{\tilde{\Delta}}  \int d^2 z   V_{1-\tilde{\Delta},  \Delta_1, \tilde{\Delta}}(z,z_1,z q)  V_{1-\tilde{\Delta},  \Delta_1, \tilde{\Delta}}(\bar{z},\bar{z}_1,0)  = \\& C_2\frac{q^{\tild} z_1^{-\Delta_1}  \bar{z}_1^{-\Delta_1}}{(1-q)^{1-\Delta_1}} \int d^2 z  \bigg{(}   
 \sum_{ n=0}^{\infty} q^n (-1)^{n-2 \tild } z^{\text{$\Delta_1 $}+n-1} (1-z)^{2 \tild -\text{$\Delta_1 $}-1}     \bar{z}^{\Delta_1-1} \times \\& \times (1-\bar{z})^{2\tild -\Delta_1-1} C^{-2 \tild -\text{$\Delta_1 $}+1} _{n}  \bigg{)}.
 \end{split}
\end{equation}
Applying the formula\footnote{For $n=0$, this formula can be written as  $ \int d^2 z  |z|^{2a}|1-z|^{2b} = \pi \frac{\gamma{(a+1) \gamma(b+1)} }{\gamma{(a+b+2)}} $, where $\gamma{(x)}= \frac{\Gamma(x)}{\Gamma(1-x)}.$}
\begin{equation} \label{integral7}
    \int d^2 z   |z|^{2a} z^n|1-z|^{2b} = \pi \frac{ \Gamma(a+n+1)\Gamma{(b+1)}  \Gamma{(-1-a-b )}} {\Gamma{ (a+n+b+2)}     \Gamma{(-a)}  \Gamma{(-b)}}  ,
\end{equation}
and integrating over $z, \bar{z}$ in (\ref{integral6}) we obtain
\begin{equation} \label{integral8}
\begin{split}
    &  C_2 q^{\tilde{\Delta}}  \int d^2 z  V_{1-\tilde{\Delta},  \Delta_1, \tilde{\Delta}}(z,z_1,z q)  V_{1-\tilde{\Delta},  \Delta_1, \tilde{\Delta}}(\bar{z},\bar{z}_1,0)  = \\& C_2 C_3\frac{q^{\tild} z_1^{-\Delta_1} \bar{z}_1^{-\Delta_1}}{(1-q)^{1-\Delta_1}} {}_2F_1(\Delta_1, \Delta_1+2\tild-1, 2\tild,q)=  z_1^{-\Delta_1} \bar{z}_1^{-\Delta_1} \mathcal{F}( \Delta_1,\tilde{\Delta},q),
 \end{split}
\end{equation}
where
\begin{equation} \label{integral9}
    C_3= \frac{\pi  \Gamma{(\Delta _1)} \Gamma (2 \tilde{\Delta }-\Delta _1) \Gamma (1-2 \tilde{\Delta })}{\Gamma (2 \tilde{\Delta }) \Gamma (1-\Delta _1) \Gamma (1-2 \tilde{\Delta }+\Delta _1)}.
\end{equation}
The dependence on $z_1, \bar{z}_1$ disappears when we divide (\ref{1ptexpansion}) by the proper normalization factor\footnote{Before, for simplicity of notations we were assuming that $\bra{\tild} \varphi_{\Delta _1}(z_1, \bar{z}_1) \ket{\tild}=1$.} $\bra{\tild} \varphi_{\Delta _1}(z_1, \bar{z}_1) \ket{\tild}=z_1^{-\Delta_1 }  \bar{z}_1^{-\Delta_1 } C_{\tild \Delta_1\tild} $. We check finally that $C_2 C_3=1$, which confirms the relation (\ref{integral1}).

To conclude this section, we highlight two observations that will be useful in the following discussions.

\hfill\break
\textbf{Observation I}
\hfill\break
Aside from an overall constant factor, the expression in (\ref{integral1}) could have been derived more straightforwardly by directly substituting the result of the three-point correlation function involving the shadow field $\tilde{\varphi}_{h^*}(z, \bar{z})$. Specifically, employing (\ref{expansiontrans}) and
\begin{equation} \label{3ptshadow}
       \bra{0}  \tilde{\varphi}_{h^*} (z, \bar{z}) \varphi_{\Delta_1}(z_1, \bar{z}_1) \varphi_{\tild} (qz, 0) \ket{0} = V_{h^*, \Delta_1, \tild} (z, z_1, z q) V_{h^*, \Delta_1, \tild} (\bar{z}, \bar{z}_1, 0),
\end{equation}
in (\ref{1ptcor5}), one can see that the result (\ref{integral1}) is obtained. This observation is crucial since it allows, in a simple way, generalizations of (\ref{integral1}) to higher-point CBs (which will be studied in the next section) and higher-spin algebras, as $\mathfrak{sl}_3$ which will be studied in section \bref{sl3blocksection}.  

\hfill\break
\textbf{Observation II} \label{observation2}
\hfill\break
The one-point CB $\mathcal{F}(\Delta_1, \tilde{\Delta},q)_{\mf{sl}_2}$ can be directly computed from the rhs of (\ref{integral1}) by performing a one-dimensional integral over $z$ within a certain  integration region. Namely, by integrating $V_{1-\tilde{\Delta}, \Delta_1, \tilde{\Delta}}(z, z_1, zq)$ over $z$ from 0 to 1, the block is recovered, up to a multiplicative constant $c_0$
 \begin{equation}  \label{obervation21}
\begin{split}
\mathcal{F}(\Delta_1, \tilde{\Delta},q)_{\mf{sl}_2} = c_0q^{\tild}  \int_{0}^1 dz V_{1-\tilde{\Delta},  \Delta_1, \tilde{\Delta}}(z,z_1,zq).
\end{split}
\end{equation}
This integral is much simpler than the original two-dimensional integral. The justification of (\ref{obervation21}) is that the integral (\ref{integral7}) factorizes into a product of two one-dimensional integrals
\begin{equation}
     \int d^2 z   |z|^{2a} z^n|1-z|^{2b} \propto \int_{0}^1 dz z^{a+n}(z-1)^b  \int_{\mathbf{C}} d\bar{z} \bar{z}^{a}(\bar{z}-1)^b,
\end{equation}
where $\mathbf{C}$ is the contour shown in Fig.\bref{fig1}. The contribution from the integral over $z$ is the relevant contribution that reproduces the block, while the integration over $\bar{z}$ gives just an overall factor. Hence, to compute the one-point CB from (\ref{integral1}), one does not need to take the two-dimensional integral but to find the proper integration contour and integrate over it. These two observations substantially simplify the computation of the CBs from the integral representation; below, we will use them in the computation for the $\mathfrak{sl}_3$ one-point torus block.

\begin{figure}
\centering
\begin{tikzpicture}
    \draw [<->,thick] (0,2) node {}    |- (4,0) node {} ;
      \node at (1, 1.7) {$\bar{z}$};
        \node at (3.2, 1.7) {$\mathbf{C}$};
      \draw (1,1.7) circle (0.2cm);
    \draw [thick] (0,-2) -- (0,0);
    \draw [thick] (-2,0) -- (0,0);
      \node at (1.7,0.6) {$1+i \epsilon $};
      \node at (1,-0.3) {$1 $};
       \fill (1,0.7) circle (2pt);
\draw[->, thick] (4,1.3) -- (3,1.3) to [out=180,in=80] (0.7,0.8) to [out=-80,in=180] (3,0.1) -- (4,0.1);  
\end{tikzpicture}
\caption{}\label{fig1}
\end{figure}

\subsubsection{Higher-point torus conformal blocks via shadow formalism} \label{1ptpw1.3}

Let us introduce the object
\begin{equation} \label{2ptblock}
  G_{\mathbf{\Delta}_2 } (\mathbf{z}_2, \bar{\mathbf{z}}_2, q) = \sum_{\tilde{\Delta}_1 \in \mc{D}} \sum_{m=0} \bra{\tilde{\Delta}_1,m} \varphi_{\Delta_1}(z_1, \bar{z}_1) \varphi_{\Delta_2}(z_2, \bar{z}_2)q^{\bL_0} \ket{m, \tild_1},
\end{equation}
which is the holomorphic contribution in $q$ to the two-point torus correlation function
\begin{equation} \label{2pt}
    \langle \varphi_{\Delta_1  } (z_1, \bar{z}_1) \varphi_{\Delta_2  } (z_2, \bar{z}_2)\rangle_{torus}. 
\end{equation}
Focusing on the $\mathfrak{sl}_2$ subsector and using standard techniques of CFT on the torus, one can decompose (\ref{2ptblock}) in the s-channel (also called the necklace channel) into global two-point torus CBs $\mc{F}_{\mathbf{\tilde{\Delta}}_2}^{\mathbf{\Delta}_2}(\mathbf{z}_2, q)$. This is achieved by inserting the identity operator (\ref{2pt1}) in (\ref{2ptblock}) between $\varphi_{\Delta_1}$ and $\varphi_{\Delta_2}$ 
\begin{equation} \label{2pt1}
   1= \sum_{\tilde{\Delta}_2 \in \mc{D}}  \sum_{m=0} \left( B_{\tilde{\Delta}_2} ^{-1}\right)^{mm}\ket{\tilde{\Delta}_2,m} \bra{m,\tilde{\Delta}_2},
\end{equation}
after that, one obtains the decomposition (for details and explicit form of $\mc{F}_{\mathbf{\tilde{\Delta}}_2}^{\mathbf{\Delta}_2}(\mathbf{z}_2, q)$ see, e.g., \cite{Alkalaev:2017bzx, Alkalaev:2022kal})
\begin{equation} \label{2pt2}
    G_{\mathbf{\Delta}_2 } (\mathbf{z}_2, \bar{\mathbf{z}}_2, q)=\sum_{\tilde{\Delta}_1, \tilde{\Delta}_2 \in \mc{D}} \tilde{C}_{\tilde{\Delta}_1 \Delta_1 \tilde{\Delta}_2 }  \tilde{C}_{\tilde{\Delta}_2 \Delta_2 \tilde{\Delta}_1} \mc{F}_{\mathbf{\tilde{\Delta}}_2}^{\mathbf{\Delta}_2}(\mathbf{z}_2, q).
\end{equation}
On the other hand, one can decompose (\ref{2ptblock}) into two-point torus conformal partial waves. This can be done in a similar way to (\ref{integral1}) by inserting twice the identity operator $\mathcal{P}$ in the summand of (\ref{2ptblock}):
\begin{equation} \label{2pt3}
\begin{split}
     \sum_{m=0} \bra{\tilde{\Delta}_1,m}  \varphi_{\Delta_1}(z_1, \bar{z}_1)  \varphi_{\Delta_2}(z_2, \bar{z}_2)q^{\bL_0}  & \ket{m, \tild_1}= \\& \sum_{m=0} \bra{\tilde{\Delta}_1,m} \mathcal{P} \varphi_{\Delta_1}(z_1, \bar{z}_1)  \mathcal{P}\varphi_{\Delta_2}(z_2, \bar{z}_2)q^{\bL_0} \ket{m, \tild_1}.
     \end{split}
\end{equation}
Substituting the explicit form of $\mathcal{P}$ in the rhs of (\ref{2pt3}), we have
\begin{equation} \label{2pt4}
\begin{split}
 & \sum_{m=0} \bra{\tilde{\Delta}_1,m}  \varphi_{\Delta_1}(z_1, \bar{z}_1)  \varphi_{\Delta_2}(z_2, \bar{z}_2)q^{\bL_0} \ket{m, \tild_1}  =\\&
\begin{split}
&
\begin{split}
   \sum_{m=0} \sum_{h_1 \in \mc{D}} \sum_{\tild_2 \in \mc{D}} &q^{\tild_1} \int d^2w_1 d^2 w_2  \bra{\tilde{\Delta}_1,m} \varphi_{h_1}(w_1, \bar{w}_1)  \ket{0}   \bra{0} \tilde{\varphi}_{h_1^*} (w_1, \bar{w}_1)\times  \\&  \times \varphi_{\Delta_1}(z_1, \bar{z}_1)  \varphi_{\tild_2} (w_2, \bar{w}_2) \ket{0}     \bra{0} \tilde{\varphi}_{\tild_2^*} (w_2, \bar{w}_2)  \varphi_{\Delta_2}(z_2, \bar{z}_2)q^{\bL_0} \ket{m, \tild_1}= \end{split} \\&  \begin{split} \sum_{\tild_2 \in \mc{D}}  C_3(\tild_1, \tild_2, \Delta_1, \Delta_2) q^{\tild_1} \int d^2w_1 d^2 w_2   & V_{\tild_1^{*}, \Delta_1, \tild_2} (w_1, z_1, w_2) V_{\tild_1^{*}, \Delta_1, \tild_2} ( \bar{w}_1, \bar{z}_1, \bar{w}_2) \times \\& \times V_{\tild_2^{*}, \Delta_2, \tild_1} (w_2, z_2, w_1 q) V_{\tild_2^{*}, \Delta_2, \tild_1} ( \bar{w}_2, \bar{z}_2, 0),
   \end{split}
\end{split}
\end{split}
\end{equation}
where in the fourth and fifth lines we used (\ref{matrixE0},\ref{3ptshadow}), and  $C_3(\tild_1, \tild_2, \Delta_1, \Delta_2)$ is a a normalization constant similar to $C_2$ of the previous section. Thus, (\ref{2ptblock}) can be written  as
\begin{equation} \label{2pt42}
\begin{split}
 & G_{\mathbf{\Delta}_2 } (\mathbf{z}_2, \bar{\mathbf{z}}_2, q)= \\& \begin{split}\sum_{\tilde{\Delta}_1, \tilde{\Delta}_2 \in \mc{D}}  C_3(\tild_1, \tild_2, \Delta_1, \Delta_2)     q^{\tild_1} \int d^2w_1 d^2 w_2   & V_{\tild_1^{*}, \Delta_1, \tild_2} (w_1, z_1, w_2) V_{\tild_1^{*}, \Delta_1, \tild_2} ( \bar{w}_1, \bar{z}_1, \bar{w}_2) \times \\& \times V_{\tild_2^{*}, \Delta_2, \tild_1} (w_2, z_2, w_1 q) V_{\tild_2^{*}, \Delta_2, \tild_1} ( \bar{w}_2, \bar{z}_2, 0).
\end{split}
\end{split}
\end{equation}
 The factor which arises in (\ref{2pt4}) 
\begin{equation} \label{2pt5}
\begin{split}
q^{\tild_1} \int d^2w_1 d^2 w_2   & V_{\tild_1^{*}, \Delta_1, \tild_2} (w_1, z_1, w_2) V_{\tild_1^{*}, \Delta_1, \tild_2} ( \bar{w}_1, \bar{z}_1, \bar{w}_2) \times \\& \times V_{\tild_2^{*}, \Delta_2, \tild_1} (w_2, z_2, w_1 q) V_{\tild_2^{*}, \Delta_2, \tild_1} ( \bar{w}_2, \bar{z}_2, 0)
\end{split}
\end{equation}
is the holomorphic contribution in $q$ of the two-point torus partial wave defined as
\begin{equation} \label{2pt6}
\begin{split}
 W^{\mathbf{\Delta}_2} _{\mathbf{\tilde{\Delta}}_2} (\mathbf{z}_2, \bar{\mathbf{z}}_2, q, \bar{q})=q^{\tild_1}  \bar{q} ^{\tild_1}\int d^2w_1 d^2 w_2   & |V_{\tild_1^{*}, \Delta_1, \tild_2} (w_1, z_1, w_2)|^2  | V_{\tild_2^{*}, \Delta_2, \tild_1} (w_2, z_2, w_1 q)|^2 .
\end{split}
\end{equation}
Equations (\ref{2pt2}, \ref{2pt42}) show the one-to-one correspondence that exists between the two-point CBs in the s-channel and the two-point torus partial waves. Therefore, we see that holomorphic two-point CBs can be expressed in terms of the holomorphic two-point conformal partial waves (see \cite{Alkalaev:2023evp}). This statement is the generalization of (\ref{integral1}). 

Further generalization to the relation between higher multi-point CBs in the necklace channel and partial waves is possible, as discussed in \cite{Alkalaev:2023evp}. One can study $n$-point torus correlation functions:
\begin{equation} \label{2pt8}
    \langle  \prod_{i=1}^n  \varphi_{\Delta_i} (z_i, \bar{z}_i)\rangle_{torus}\;,
\end{equation}
and define the $n$-point torus partial waves
\begin{equation} \label{2pt9}
    W_{\mathbf{\tilde {\Delta}}_n } ^{\mathbf{\Delta}_n} (\mathbf{z}_n,\bar{\mathbf{z}}_n,q, \bar{q}) = q^{\tild_1} \bar{q}^{\tild_1}  \int_{\mathbb{R}^n}   \left(   \prod_{j=1}^n  d^2w_j | V_{ \tild_j^*, \Delta_j, \tild_{j+1}} (w_j, z_j, w_{j+1}) |^2\right),
\end{equation}
with the identifications $\tild _{n+1}=\tild_1$, $w_{n+1}=q w_1$. By inserting the resolution of identity operator $\mathcal{P}$ $n$-times between the external fields $\varphi_{\Delta_i}(z_i, \bar{z}_i)$ in the expression for (\ref{2pt8}), in a similar way to (\ref{2pt3}), we deduce that (\ref{2pt8}) can be expressed as a sum of $n$-point partial waves. This leads to the conclusion that these partial waves are related in a nontrivial way to the $n$-point torus CBs in the necklace channel.

\section{$\mathfrak{sl}_3$ global conformal blocks   } \label{sl3blocksection}
\label{prel} 

\subsection{Preliminaries: $\varW_3$ Conformal Field Theory} 
\label{theory1} 

The $\mc{W}3$ CFT is an extension of the Virasoro CFT. The symmetry of the $\varW_3$ CFT is generated by the energy-momentum tensor $\mathbf{T}(z)$ and an additional spin-3 current $\mathbf{W}(z)$. The Laurent series expansion of $\mathbf{W}(z)$ reads:
\begin{equation}
\bW(z)= \sum_{n=-\infty}^{\infty} \frac{\bW_n}{z^{n+3}}.
\end{equation}
The modes $\bL_n$ and $\bW_m$ generate the $\varW_3$ algebra, which reads
\begin{equation}
 \begin{split}
\left[\bL_n, \bL_m\right]		& =(n-m) \bL_{n+m}+\dps \frac{c}{12}(n^3-n) \delta_{n+m,0} \; , \\ 
\left[\bL_n, \bW_m\right]		& =(2n-m) \bW_{n+m} \; , \\ 
\left[\bW_n, \bW_m\right]		& = \frac{c}{3\cdot5!}(n^2-1)(n^2-4)n \delta_{n+m,0}+\frac{16}{22+5c}(n-m) \mathbf{\Lambda}_{n+m} + \\ 
						& \qquad + \frac{(n-m)}{30}\left( 2m^2 +2n^2 - m n -8  \right)\bL_{n+m} \; ,
\end{split}
\end{equation}
where
\begin{equation}
\mathbf{\Lambda}_{m} = \sum_{p \leq - 2} \bL_{p} \bL_{m-p} + \sum_{p \geq - 1} \bL_{m-p} \bL_{p} - \frac{3(m+2)(m+3)}{10} \bL_{m} \; .
\label{nl} 
\end{equation}
The Virasoro algebra (\ref{sl2gbs2}) is a subalgebra of $\varW_3$ algebra. In the limit $c\rightarrow \infty$ these commutation relations reduce to the ones of the $\sl_3$ algebra, generated by
\begin{equation} 
\{ \bL_{-1}, \bL_0, \bL_1, \bW_{-1}, \bW_1, \bW_0, \bW_{-2}, \bW_2 \} \; ,
\label{sl3generators} 
\end{equation}
that satisfy
\begin{equation}
\begin{split}
[\bL_n, \bL_m] 		& = (n-m) \bL_{n+m} \; , \\
[\bL_n, \bW_m] 	& = (2n-m)\bW_{n+m} \; ,  \\ 
[\bW_n, \bW_m] 	& = (n-m)\Big(\frac{1}{15}(n+m+2)(n+m+3)-\frac{1}{6}(n+2)(m+2)\Big) \bL_{n+m} \; . 
\end{split}
\label{eqn:W3.1} 
\end{equation}
\paragraph{$\varW_3$ primary fields.} 
We denote the $\varW_3$ primary fields characterized by the vector $j$ as 
\begin{equation} \label{sl3primf}
    \phi_{j}(z,\bar{z}),
\end{equation}
where $j$ belongs to root space of $\mf{sl}_3$, generally written in the form (\ref{vofsl3}). Alternatively, $\phi_{j}(z,\bar{z})$, can be characterized by two parameters $h_{j},q_{j}$ corresponding to the conformal dimension and $\varW_3$ charge of $\phi_{j}$, given by
\begin{equation}
h_{j} = \frac{1}{2}(\alpha_{j}, \, 2Q-\alpha_j ) \, , \qquad q_{j} = i\sqrt{\frac{48}{22 + 5c}}\prod_{i=1}^{3}(e_i, \, \alpha_j - Q) \; .
\label{dim} 
\end{equation}
Here
$Q = (b + \frac{1}{b})(w_1 + w_2)$, $e_i$ are the weights of the fundamental representation
\begin{equation}  
e_1= w_1\, , \quad  e_2= w_2-w_1\, , \quad e_3= -w_2 \; ,
\label{voffundamentalr}
\end{equation}
and $\alpha_j$ is also a vector on the root space of $\mf{sl}_3$. In the large central charge limit, $\alpha_j$ is given by
\begin{equation}
\alpha_j =-bj,
\label{alpha2} 
\end{equation}
for which the conformal dimension and the charge assume the values
\begin{equation}
h_{j}= -r-s, \qquad q_{j} = \frac{i}{3}\sqrt{\frac{2}{5}}(s-r) \; .
\label{dimd} 
\end{equation}
The $\varW_3$ primary fields satisfy (similarly to (\ref{sl2gbs6})) the commutation relations
\begin{equation}
\begin{split}
&[ \bL_{n} , \phi_j (z, \bar{z})] = z^n \left(   z \partial_{z}+h_j (n+1) \right) \phi_j(z, \bar{z}),\\&
[ \bW_n, \phi_j(z, \bar{z})]= z^n\left( \frac{q_j}{2}(n+2)(n+1) +(n+2)z \hat{W}_{-1}+z^2  \hat{W}_{-2}  \right)\phi_j(z, \bar{z}),
\end{split}
\end{equation}
 where, in contrast to $\bL_n$, the commutation relations with $\bW_n$ incorporate two additional operators, $\hat{W}_{-1}$ and $\hat{W}_{-2}$ \cite{Watts:1994zq}, which cannot be expressed as differential operators in terms of the variable $z$.

A $\varW_3$ \emph{highest-weight vector} $\ket{j}$ given by
\begin{equation} \label{w3hwv}
\ket{j} = \lim_{z \rightarrow 0}  \phi_{j}(z, \bar{z}) \ket{0}\; , 
\end{equation}
satisfies the conditions
\begin{gather} \label{w3hwvcon}
\bL_0 \ket{j} = h_{j} \ket{j} \; , \qquad \bW_0 \ket{j} = q_{j} \ket{j} \; , \\
\bL_{n}\ket{j} = \bW_{n}\ket{j} = 0 \; , \qquad n > 0 \; .
\end{gather}
The $\varW_3$ \emph{module} associated with this highest-weight vector is spanned by a basis of descendant states
\begin{equation} 
\mathcal{L}_{-I}   \ket{j}  =  \bL_{-i_1} \dots \bL_{-i_m} 
   \bW_{-k_1} \dots \bW_{-k_n} \ket{j} \, , \quad I= \{i_1, \dots, i_m; k_1,\dots, k_n \} \; , \\
\label{w3module} 
\end{equation}
with 
\be
1 \le i_1 \le \dots \le i_m \, , \quad  1 \le k_1 \le \dots \le k_n \; .
\ee
The sum 
\begin{equation}
    \sum_{i_a, k_b \in I} i_a + k_b
\end{equation}
is called \textit{level} of the state $\mathcal{L}_{-I}   \ket{j} $. We will focus only on the $\mf{sl}_3$ \textit{module} (which is a subspace of (\ref{w3module})) spanned by the basis of states 
\begin{equation} \label{sl3mod}
\ket{N,j}=    \left(\bW_{-2}\right)^{n_3}\left( \bW_{-1} \right) ^{n_2}\left(  \bL_{-1} \right)^{n_1}\ket{j}, \quad \text{ where  $N=(n_1, n_2, n_3) \in \mathbb{I}$}.
\end{equation}
where we define $\mathbb{I}$ as
\begin{equation}
   \mathbb{I}= \{   (n_1, n_2, n_3) :   (n_1, n_2, n_3) \in \text{non-negative integers}   \}.
\end{equation}
The level of the states (\ref{sl3mod}) is given by
\begin{equation}
    |N| =n_1+n_2+2 n_3.
\end{equation}
\subsection{$\mathfrak{sl}_3$ global one-point torus conformal blocks} 
\label{1ptw3} 
The holomorphic $\mathfrak{sl}_3$ global one-point CB $\mathcal{F}(j_1, j, q)_{\mf{sl}_3}$\footnote{For further discussions related to the $\mathcal{W}_3$ one-point CBs see also \cite{Chang:1991ht, He:2012bi, Hadasz:2009db}.} is defined as
\begin{equation} 
\mathcal{F}(j_1, j , q)  = \frac{1}{ \bra{j} \phi_{j_1}(z_1,\bar{z}_1) \ket{j}} \sum_{\substack{M \in \mathbb{I} \\ |M|=|N|}} \sum_{ N \in \mathbb{I}}\bra{j, M}  \phi_{  j_1}(z_1, \bar{z}_1)  q^{\bL_0} \ket{N, j} \left(  B_j^{-1} \right)^{MN} \; , 
\label{1pt} 
\end{equation}
where the states $\ket{ N,j} $ belong to the $\mf{sl}_3$ module~\eqref{sl3mod}, and $\left( B_j^{-1} \right)^{MN}$ is the inverse of the Shapavalov matrix 
\begin{equation}
B_j ^{MN} = \braket{j, M |  N,j}  \; . 
\end{equation}
Unlike the $\mf{sl}_2$ one-point torus block, to date, there is no known exact expression for (\ref{1pt}). One reason for this is the lack of a general expression for the matrix elements
\begin{equation} \label{sl3me}
\bra{j, M} \phi_{j_1}(z_1, \bar{z}_1) \ket{ N,j} .\;  
\end{equation}
In fact, for a general $\phi_{j_1}$ (\ref{sl3me}) is not uniquely defined. In this work, we will concentrate on the case when $\phi_{j_1}$ is a degenerate field at the first level, satisfying~\eqref{alpha4}. This condition is fulfilled when $j_1$ consists of a single component. In the parameterization
\begin{equation}
    j= r w_1 +s w_2, \quad j_1= 3a w_1,
\end{equation}
a perturbative expression of (\ref{1pt}) up to the second level (up to $q^2$) was computed in \cite{Belavin:2023orw}, and looks as follows
\begin{equation} 
\begin{split}
\mathcal{F}(j_1, j,q)_{\mathfrak{sl}_3} 	& = 1 + \bigg( \frac{2rs - a^2(r+s) - a(r + s)}{rs}\bigg)q \, + \\
            & \quad + \frac{1}{162}\bigg( \frac{3a(3a+6)(3a+3)(3a-3)}{s - 1} \, + \\
            & \quad + \frac{\big(3a(3a+3) - 18r\big)\big(3a(3a+3) - 36(r - 1)\big)}{r(r - 1)} \, + \\
            & \quad - \frac{3a(3a+3)\big(3a(3a+3)(r - 1) + 36(r + 1)\big)}{(r + 1)s} \, + \\
            & \quad + \frac{6a(3a+3)(3a-3r)(3a+3r + 3)}{(r + 1)r(r + s + 1)}\bigg) q^2 + \dots \; .
\end{split}   
\label{sl31ptper} 
\end{equation}
The computation of higher-level contributions to (\ref{sl31ptper}) is a very challenging task, specially when one computes them directly from (\ref{sl3me}). The expression (\ref{sl31ptper}) was verified by two other methods, using the AGT relation and the Wilson lines interpretation of CBs in AdS$_3$. In this paper, we will provide an exact expression for (\ref{1pt}) using the Shadow formalism.

\section{$\mathfrak{sl}_3$ conformal blocks via shadow formalism} \label{geralsl3shf}
This section aims to extend the shadow formalism theory from sections (\bref{basiccon1}, \bref{1ptpw1}) to the $\mf{sl}_3$ case, enabling us to compute the $\mf{sl}_3$ global one-point torus block (\ref{1pt}). For $\mf{sl}_3$ global four-point sphere CBs, a similar approach to the shadow formalism was presented in \cite{Fateev:2011qa}. We will reformulate this approach using the language of shadow formalism to apply it to torus topology. We begin by reviewing the theory of $\mf{sl}_3$ invariant functions.

\subsection{$\mf{sl}_3$ invariant functions} \label{subsecsl3invf}
The $\mathfrak{sl}_3$ algebra can be represented in the Chevalley basis, which includes two Cartan elements ($h^1, h^2$), along with creation and annihilation generators ($e^1, e^2, e^3$) and ($f^1, f^2, f^3$), respectively. These elements satisfy the following relations
\begin{equation} \label{sl3algebra}
\begin{split}
&[f^1, f^2]=-f^3, \quad [e^1, e^2]=e^3,    \quad [e^1, f^1]= h^1,\\&
[e^2,f^2] = h^2, \quad [e^3,f^3]=h^1+h^2, \quad [e^1,f^3]= -f^2,\\ &
[e^2,f^3]= f^1,  \quad [e^3,f^1]=-e^2,    \quad  [e^3,f^2]=e^1,\\
\end{split}
\end{equation}
with all other commutators being zero. The generators in (\ref{sl3algebra}) can be expressed as linear combinations of the generators in (\ref{eqn:W3.1}). The generators (\ref{sl2tranfor}) correspond to the $\mf{sl}_2$ transformation for the variable $z$ (representing the physical coordinates of the fields). To represent the generators of $\mf{sl}_3$ in terms of differential operators, we introduce three-component isospin variables $Z$
\begin{equation}  \label{sl3sec2}
    Z= (w, x, y).
\end{equation}
For a given vector $j=r w_1 + sw_2$, the generators (\ref{sl3algebra}) can be constructed as differential operators acting on monomials of the form $ x ^a    y ^b w ^c$ for $a+c\le r$, $b\le s$, as follows
\begin{equation} \label{sl3sec3}
\begin{split}
   &  D_{(j,Z)} (h^1)= 2 x \partial_x+r-y \partial_y +w \partial _w ,\\&
    D_{(j,z)}(h^2) = 2 y \partial_y+s-x \partial_x+w \partial _w, \\&
    D_{(j,Z)} (e^1) =x^2 \partial_x+r x +(w- x y )\partial_y+ x w \partial_w, \\&
    D_{(j,Z)}(e^2)= y^2 \partial_y+sy-w\partial_x,\\&
    D_{(j,Z)}(e^3) = w^2 \partial_w+s(w-x y) +r w+xw \partial_x+y (w- xy)\partial _y,\\&
    \partial_{(j,Z)}(f^1)=-\partial_x,\\&
    D_{(j,Z)} (f^2)= -\partial_y-x \partial_w,\\&
    D_{(j,Z)}(f^3)=-\partial_w.
\end{split}
\end{equation}
To construct $\mf{sl}_3$ invariant functions, we define the following notations:
\begin{equation} \label{sl3sec3}
\begin{split}
& j_i=r_i w_1 + s_i w_2 = (r_i, s_i),\quad  j_i^w = (s_i,r_i), \quad j_i^* = (2-s_i, 2-r_i),\\
& Z_i= (w_i, x_i, y_i),\\
    &\rho_{ij}=y_i \left(x_i-x_j\right)-\left(w_i-w_j\right),\\&
    \chi_{ijk}=y_i w_j-w_i y_j+y_i y_j \left(x_i-x_j\right)-y_i w_k+w_i y_k+\\& +y_i y_k \left(x_k-x_i\right)-w_j y_k+y_j w_k+y_j y_k \left(x_j-x_k\right),\\&
    \sigma_{ijk}=x_i w_j-w_i x_j-x_i w_k+w_i x_k-w_j x_k+x_j w_k,
    \end{split}
\end{equation}
where $j_i^*$ and $j_i^w$ correspond to the maximal Weyl transformation and Dynkin automorphism of the spin $j_i$, respectively. An $\mathfrak{sl}_3$ \textit{invariant $n$-point function} associated with $n$ spins $j_1, j_2, ..., j_n$ is a function $\xi(j_i | Z_i)$ such that
\begin{equation} \label{invariantc}
    \left(  \sum_{i=1}^n D_{(j_i,Z_i)}(t^a) \right) \xi(j_i|Z_i) =0,\quad \text{for any $t^a \in \{ h^i, e^i, f^i \}$}.
\end{equation}
The function $\xi$ will satisfy additional equations if some representations are labeled by only one fundamental weight: 
\begin{equation} \label{sl3sec4}
\begin{split}
 &   \text{if for some $k$ } \quad (j,w_k) =0 , \quad \text{then}, \\&
 d^{(k)}_{Z_k} \xi(j_i|Z_i) =0,
\end{split}
\end{equation}
where 
\begin{equation} \label{sl3sec5}
    d^{(1)}_Z= \partial_x+y \partial_w, \quad d^{(2)}_Z=\partial_y.
\end{equation}
For two spins $j_1$ and $j_2$, the $\mf{sl}_3$ invariant two-point function is given by
\begin{equation} \label{sl32ptinv}
    \xi (j_1, j_2|Z_1,Z_2) = \begin{cases}\rho_{21}^{-r_1}\rho_{12}^{-s_1}, \quad \text{for $j_1= j_2^{w}$} ,\\0 , \quad \text{otherwise} . \end{cases}
\end{equation}
For general representations $j_1, j_2, j_3$, the $\mf{sl}_3$ invariant three-point function is not uniquely determined by (\ref{invariantc}). It can be specified up to a general function $g(\theta_{123})$, where $\theta_{123}=\frac{\rho_{12} \rho_{23} \rho_{31}}{ \rho_{21} \rho_{32}\rho_{13}}$. In the case of interest, where one of the representations has only one component, say $j_2 = (0,s_2)$, and the others, $j_1 = (r_1,s_1)$ and $j_3 = (r_3,s_3)$, have two components, the $\mathfrak{sl}_3$ invariant three-point function $\xi$ is given by
\begin{equation}\label{invafun}
    \xi(j_1, j_2, j_3|Z_1, Z_2,Z_3)= \chi_{123}^{-J}\rho_{21}^{-J-r_1+s_3} \rho_{23}^{-J-r_3+s_1} \rho_{13}^{J-s_1} \rho_{31}^{J-s_3}, 
\end{equation}
where  $J=(w_2-w_1)\cdot (j_1+j_2+j_3) $, in our case $J= \frac{1}{3}(s_1+s_2+s_3-r_1-r_3)$. Two and three-point $\mf{sl}_3$ invariant functions (\ref{sl32ptinv}, \ref{invafun}) are the analogs of the two-point function $(z_1-z_2)^{-2\Delta}$ and three-point function (\ref{3ptcoord}).

\subsection{$\mf{sl}_3$ shadow formalism} \label{sl3shadowformconst}
To extend the Shadow formalism to the $\mf{sl}_3$ algebra, we introduce $\mf{sl}_3$ fields $\Phi_j(Z, \bar{Z},g)$. These fields depend on isospin variables $Z$, rather than on the two-dimensional complex coordinates ($z, \bar{z}$), and are labeled by an element $g \in SL_3$ and a representation $j$. Detailed expressions for these fields are provided in \cite{Fateev:2011qa} and Appendix \bref{apsl3fields}. They behave as primary fields under $\mf{sl}_3$ transformations of $g$, satisfying
\begin{equation} \label{sl3trans}
    \Phi_{j}(Z, \bar{Z}, (1+\epsilon t^a)g)= (1 +  \epsilon D_{(j, Z)}(t^a)   ) \Phi_{j}(Z, \bar{Z},g).
\end{equation}
where $D_{(j, Z)}(t^a)$ are from (\ref{sl3sec3}). In the large $c$ limit, the correlation functions of primary fields $\phi_{j_i}(z_i, \bar{z_i})$ in $\mf{sl}_3$ conformal Toda theory become \cite{Fateev:2007ab}
\begin{equation}
\lim_{c \to \infty}   \langle  \phi_{j_1}(z_1, \bar{z}_1)...\phi_{j_n}(z_n, \bar{z}_n)  \rangle  = \langle  \Phi_{j_1}(\vec{z}_1, \vec{\bar{z}}_1)...\Phi_{j_n}(\vec{z}_n, \vec{\bar{z}}_n)    \rangle ,
\end{equation}
where we define 
\begin{equation} \label{defofPhiprod1}
     \langle  \Phi_{j_1}(Z_1, \bar{Z}_1)...\Phi_{j_n}(Z_n, \bar{Z}_n)    \rangle:= \int_{SL_3} dg   \Phi_{j_1}(Z_1, \bar{Z}_1,g)...\Phi_{j_n}(Z_n, \bar{Z}_n,g)   ,
\end{equation}
and
\begin{equation}
    \vec{z}:= (\frac{z^2}{2},z,z).
\end{equation}
Clearly, the rhs of (\ref{defofPhiprod1}) is invariant under (\ref{sl3trans}), hence the ``correlation function''
\begin{equation} \label{nptofPhi}
     \langle  \Phi_{j_1}(Z_1, \bar{Z}_1)...\Phi_{j_n}(Z_n, \bar{Z}_n)    \rangle
\end{equation}
is an $\mf{sl}_3$ invariant $n$-point function. This implies that (\ref{sl32ptinv}) is interpreted as the two-point correlation function of $\mf{sl}_3$ fields
\begin{equation}
    \langle \Phi_{j_1}(Z_1, \bar{Z}_1) \Phi_{j_2}(Z_2, \bar{Z}_2) \rangle = |\xi (j_1, j_2|Z_1,Z_2)|^2.
\end{equation}
Similarly for the $\mf{sl}_3$ three-point function (\ref{invafun}). Since the integration over the group in (\ref{defofPhiprod1}) does not interfere with the calculation procedure we will apply, we will omit the label $g$ in $\Phi_{j}$ and use the following notation
\begin{equation}
   \Phi_{j}(Z, \bar{Z}):= \Phi_{j}(Z, \bar{Z},g),
\end{equation}
keeping in mind that the correlation functions of $\Phi_{j}(Z, \bar{Z})$ are given by (\ref{defofPhiprod1}) by definition. 

The construction of the shadow formalism for $\mf{sl}_3$ starts with the construction of the shadow field. We define the shadow field as\footnote{Notice that in (\ref{shadfield}), we set the normalization constant of the type $\mathcal{N}_{\Delta}$ equal to 1. This is just a matter of convention.}
\begin{equation} \label{shadfield}
    \Phi_{j^*}(Z, \bar{Z})= \int d ^2Z'  K_{j^*} (Z,Z') \Phi_j(Z', \bar{Z}'),
\end{equation}
where $ d^2 Z'=dZ' d \bar{Z}'$, and the kernel $K_{j^*}(Z, Z')$ is given by the two-point function
\begin{equation} \label{sl3sec8}
    K_{j^*}(Z, Z')= \langle \Phi_{j^*} (Z, \bar{Z}) \Phi_{j^{*w}} (Z', \bar{Z}')   \rangle  = |\xi(j^*, j^{*w}|Z,Z'  )|^2.
\end{equation}
Analogously to (\ref{sl2gbs9}, \ref{sl2gbs11}), we define the operators 
\begin{equation} \label{sl3sec10}
     P_j =  \int d^2 Z \Phi_{j^w}(Z, \bar{Z}) \ket{0} \bra{0} \Phi_{j^{*}} (Z, \bar{Z}),
\end{equation}
and 
\begin{equation} \label{sl3sec11}
    \mathbf{P} = \sum_{j} P_j.
\end{equation}
Let us verify that the operator $\mathbf{P}$ acts as the identity operator, namely
\begin{equation} \label{iprosl3}
\begin{split} 
& \mathbf{P} \Phi_{j_1}(Z_1, \bar{Z_1}) \ket{0} =   \Phi_{j_1}(Z_1, \bar{Z}_1) \ket{0}, \\&
\bra{0} \Phi_{j_1}(Z_1, \bar{Z}_1) \mathbf{P}= \bra{0} \Phi_{j_1}(Z_1, \bar{Z}_1) .
\end{split}
\end{equation}
This property (\ref{iprosl3}) will play an important role, similar to (\ref{iprosl2}) in the $\mathfrak{sl}_2$ case. Let us verify (\ref{iprosl3}) using two different methods. The first check follows directly. Writing explicitly $\mathbf{P}$, we have
\begin{equation} \label{sl3sec12}
    \mathbf{P} \Phi_{j_1}(Z_1, \bar{Z}_1) \ket{0}= \sum_j \int d^2 Z    \Phi_{j^w} (Z, \bar{Z}) \ket{0} \bra{0} \Phi_{j^*} (Z, \bar{Z}) \Phi_{j_1} (Z_1, \bar{Z}_1) \ket{0}.
\end{equation}
By using
\begin{equation} \label{sl3sec13}
    \bra{0} \Phi_{j^*} (Z, \bar{Z}) \Phi_{j_1} (Z_1, \bar{Z}_1) \ket{0} = \delta_{j^*, j_1^w} | \xi (j^*, j_1| Z, Z_1)|^2=  \delta_{j^*, j_1^w} | \xi (j_1, j^*| Z_1, Z)|^2,
\end{equation}
 (\ref{sl3sec12}) becomes
\begin{equation} \label{sl3sec14}
     \mathbf{P} \Phi_{j_1}(Z_1, \bar{Z}_1) \ket{0}=  \int d^2 Z    \Phi_{j_1^*} (Z, \bar{Z}) \ket{0} | \xi (j_1, j_1^w| Z_1, Z)|^2 = \Phi_{j_1}(Z_1, \bar{Z}_1)\ket{0},
\end{equation}
where in the last equality we applied the definition (\ref{shadfield}), and the fact that the field $\Phi_{j_1}$ can be written as the shadow of $\Phi_{j_1^*}$. The same procedure can be applied to show $\bra{0} \Phi_{j_1}(Z_1, \bar{Z}_1) \mathbf{P}= \bra{0} \Phi_{j_1}(Z_1, \bar{Z}_1)$. A more rigorous proof is as follows. If $\mathbf{P}$ acts as the identity operator, namely, $\bra{0} \Phi_{j_1}(Z_0, \bar{Z}_0) \mathbf{P}= \bra{0} \Phi_{j_1}(Z_0, \bar{Z}_0)$, then
\begin{equation} \label{sl3sec15}
    \bra{0} \Phi_{j_1}(Z_0, \bar{Z}_0) \mathbf{P} \Phi_{j_1^w} (Z_3, \bar{Z}_3) \ket{0}= |\xi(j_1, j_1^{w}|Z_0, Z_3)   |^2  =|\rho_{30}|^{-2r_1} |\rho_{03}|^{-2s_1}.
\end{equation}
Let us check (\ref{sl3sec15}). Writing explicitly $\mathbf{P}$ we have
\begin{equation} \label{sl3sec16}
\begin{split}
&    \bra{0} \Phi_{j_1}(Z_0, \bar{Z}_0) \mathbf{P} \Phi_{j_1^w} (Z_3, \bar{Z}_3) \ket{0}= \\& \bra{0} \Phi_{j_1} (Z_0, \bar{Z}_0)  \sum_{j} \int d^2Z_1 \Phi_{j^w}(Z_1, \bar{Z}_1) \ket{0} \bra{0} \Phi_{j^*} (Z_1, \bar{Z}_1) \Phi_{j_1^w}(Z_3, \bar{Z}_3) \ket{0}.
\end{split}
\end{equation}
By inserting $\Phi_{j^*}$ from (\ref{shadfield}), and denoting the expression by $I_{j_1}$, we have
\begin{equation} \label{sl3sec17}
\begin{split}
&
I_{j_1}=   \bra{0} \Phi_{j_1}(Z_0, \bar{Z}_0) \mathbf{P} \Phi_{j_1^w} (Z_3, \bar{Z}_3) \ket{0}= \\&  \begin{split}  \bra{0} \Phi_{j_1} (Z_0, \bar{Z}_0)  \sum_{j} \int d^2Z_1 
 & \Phi_{j^w}(Z_1, \bar{Z}_1) \ket{0} \times \\& \times \bra{0}  \int d^2 Z_2 \Phi_{j} (Z_2, \bar{Z}_2) |\xi(j^*, j^{*w}| Z_1, Z_2)|^2 \Phi_{j_1^w}(Z_3, \bar{Z}_3) \ket{0}= \end{split} \\& \int d^2 Z_1  d^2 Z_2 |\xi (j_1, j_1^w|Z_0,Z_1)|^2 |\xi (j_1^*, j_1^{*w}|Z_1, Z_2)|^2 |\xi (j_1, j_1^{w}|Z_2, Z_3)|^2.
\end{split}
\end{equation}
One can show that the last integral is given by the rhs of (\ref{sl3sec15}). Indeed, by writing this integral in terms of the components of the isospin variables, we have
\begin{equation}  \label{sl3sec18}
\begin{split}
   &I_{j_1}= \int d^2 Z_1  d^2 Z_2 |\xi (j_1, j_1^w|Z_0,Z_1)|^2 |\xi (j_1^*, j_1^{*w}|Z_1, Z_2)|^2 |\xi (j_1, j_1^{w}|Z_2, Z_3)|^2=\\& \int d^2 w_1 d^2 x_1 d^2 y_1  d^2 w_2 d^2 x_2 d^2 y_2  
   |y_1 x_{10}-w_{10}|^{-2 r_1}   |y_0 x_{01}-w_{01}|^{-2 s_1}   |y_2 x_{21}-w_{21}|^{-2(2-s_1)} \times \\& \times |y_1 x_{12}-w_{12}|^{-2(2- r_1)}  |y_3 x_{32}-w_{32}|^{-2 r_1}   |y_2 x_{23}-w_{23}|^{-2 s_1} ,
    \end{split}
\end{equation}
where $x_{ij}=x_i-x_j$, $w_{ij}=w_i-w_j$. Performing the integration over $y_1$, we obtain a delta function
\begin{equation}  \label{sl3sec19}
\begin{split}
    I_{j_1}= C_1 \int d^2 w_1 d^2 x_1 d^2 w_2 d^2 x_2 d^2 y_2 & |x_{10}|^{-2r_1} |x_{12}|^{-2(2-r_1)}   \delta^2\left(\frac{w_{10}}{x_{10}}-\frac{w_{12}}{x_{12}}\right)  |y_0 x_{01}-w_{01}|^{-2 s_1} \times \\&  |y_2 x_{21}-w_{21}|^{-2(2-s_1)}    |y_3 x_{32}-w_{32}|^{-2 r_1}   |y_2 x_{23}-w_{23}|^{-2 s_1} ,
    \end{split}
\end{equation}
where $C_i$ ($i=1,2,3$) are constants present in the delta function formula. Performing the integration over $w_1$ and $y_2$, we obtain
\begin{equation}  \label{sl3sec20}
\begin{split}
    I_{j_1}=  C_{2} \int  d^2 x_1 d^2 w_2 d^2 x_2 & |x_{10}|^{-2(r_1+s_1-1)} |x_{12}|^{-2(3-r_1-s_1)}|x_{02}|^{-2 (1-s_1)} |x_{23}|^{-2 s_1} \times \\&  |y_0 x_{02}-w_{02}|^{-2 s_1}  |y_3 x_{32}-w_{32}|^{-2 r_1} \delta^2\left(\frac{w_{20}}{x_{20}}-\frac{w_{23}}{x_{23}}\right)  .
    \end{split}
\end{equation}
Performing the integration over $w_2$, we get
\begin{equation} \label{sl3sec21}
    \begin{split}
    I_{j_1}=  C_{3} \int  d^2 x_1  d^2 x_2 & |x_{10}|^{-2(r_1+s_1-1)} |x_{12}|^{-2(3-r_1-s_1)} 
 |x_{03}|^{-2( -s_1-r_1+1)}|x_{23}|^{-2( s_1+r_1-1)} \times \\&  |y_0 x_{03}-w_{03}|^{-2 s_1}  |y_3 x_{03}-w_{03}|^{-2 r_1}   ,
    \end{split}
\end{equation}
and then, after integrating over $x_1$ and $x_2$, we finally obtain
\begin{equation} \label{sl3sec22}
    I_{j_1}= C  |\rho_{30}|^{-2r_1}  |\rho_{03}|^{-2s_1}.
\end{equation}
This proves (\ref{sl3sec15}) (up to the normalization constant $C$). Thus, the above proof shows that $\mathbf{P}$ plays the role of the resolution of the identity operator. Since there is such a resolution of identity in $\mathfrak{sl}_3$, one can proceed as in the $\mathfrak{sl}_2$ case.
\subsection{$\mf{sl}_3$ global four-point sphere conformal blocks via shadow formalism} \label{sl34ptsphereCB}
In $\varW_3$ CFT, analogous to (\ref{spblock1}), one can decompose the four-point correlation function involving external fields free of multiplicities into CBs. On the other hand, one can use shadow formalism to obtain an integral representation of the conformal block. For this, one needs to compute the four-point correlation function of $\mf{sl}_3$ fields
\begin{equation} \label{sl34ptsphere1}
    \langle \Phi_{j_1} (\vec{z}_1, \vec{\bar{z}}_1 ) \Phi_{j_2} (\vec{z}_2, \vec{\bar{z}}_2) \Phi_{j_3} (\vec{z}_3, \vec{\bar{z}}_3) \Phi_{j_4} (\vec{z}_4, \vec{\bar{z}}_4) \rangle.
\end{equation}
To avoid the problem of multiplicities, one chooses
\begin{equation} \label{thefourspins}
    j_1=(r_1, s_1), j_2 = (0,s_2), j_3= (0,s_3), j_4= (r_4, s_4).
\end{equation}
Then, one inserts the resolution of identity operator $\mathbf{P}$ as follows
\begin{equation}
\begin{split} \label{sl34ptsphere2}
  &    \langle \Phi_{j_1} (\vec{z}_1, \vec{\bar{z}}_1 ) \Phi_{j_2} (\vec{z}_2, \vec{\bar{z}}_2)  \mathbf{P} \Phi_{j_3} (\vec{z}_3, \vec{\bar{z}}_3) \Phi_{j_4} (\vec{z}_4, \vec{\bar{z}}_4) \rangle =    \\&    \langle \Phi_{j_1} (\vec{z}_1, \vec{\bar{z}}_1 ) \Phi_{j_2} (\vec{z}_2, \vec{\bar{z}}_2)  \sum_{j} \int d^2 Z \Phi_{j^w}(Z, \bar{Z}) \ket{0} \bra{0} \Phi_{j^{*}} (Z, \bar{Z}) \Phi_{j_3} (\vec{z}_3, \vec{\bar{z}}_3) \Phi_{j_4} (\vec{z}_4, \vec{\bar{z}}_4) \rangle.
    \end{split}
\end{equation}
The result of this procedure is that one decomposes (\ref{sl34ptsphere1}) in terms of the following object
\begin{equation} \label{sl34ptsphere3}
\begin{split}
&  \int d^2 Z   \langle \Phi_{j_1} (\vec{z}_1, \vec{\bar{z}}_1 ) \Phi_{j_2} (\vec{z}_2, \vec{\bar{z}}_2)    \Phi_{j^w}(Z, \bar{Z}) \ket{0} \bra{0} \tilde{\Phi}_{j^{*}} (Z, \bar{Z}) \Phi_{j_3} (\vec{z}_3, \vec{\bar{z}}_3) \Phi_{j_4} (\vec{z}_4, \vec{\bar{z}}_4) \rangle = \\& \int d^2 Z  |\xi (j_1, j_2, j^w| \vec{z}_1, \vec{z}_2, Z)|^2
  |\xi (j^*, j_3, j_4| Z, \vec{z}_3, \vec{z}_4  )|^2.
  \end{split}
\end{equation}
To simplify the above integral and identify this object with the $\mf{sl}_3$ global four-point sphere CB $\mc{F}^s( \mathbf{j}_4, j, \mathbf{ z}_4)_{\mf{sl}_3}$ (where $\mathbf{j}_4=j_1, j_2, j_3, j_4$), one needs to find the proper integration contour $C^s_{\mf{sl}_3}$ over $Z$ (this is given in appendix B.1 of \cite{Fateev:2011qa}). After this simplification, one obtains the CB from (\ref{sl34ptsphere3}) as follows
\begin{equation} \label{sl34ptsphere4}
    \mc{F}^s( \mathbf{j}_4, j, \mathbf{ z}_4)_{\mf{sl}_3}=\mc{N}^s \int_{C^s_{\mf{sl}_3}} d Z  \xi (j_1, j_2, j| \vec{z}_1, \vec{z}_2, Z)
  \xi (j^{*w}, j_3, j_4| Z, \vec{z}_3, \vec{z}_4  ),
\end{equation}
where $\mc{N}^s$ is a normalization constant chosen properly to have the correct asymptotic behavior, and the function $\xi(j_1, j_2, j_3|Z_1, Z_2, Z_3)$ is given by (\ref{invafun}). The above result (\ref{sl34ptsphere4}) reproduces the known result of the $\mf{sl}_3$ global four-point CB \cite{Fateev:2011qa}, which has also been obtained from the AdS$_3$ holographic perspective \cite{Besken:2016ooo}. 

\subsection{$\mf{sl}_3$ global one-point torus conformal blocks via shadow formalism} \label{sl3globalonepointtorus}

In this section, we study the $\mf{sl}_3$ global one-point torus block using the shadow formalism developed in section \bref{sl3shadowformconst} and provide an expression for the CB. We start by presenting the result. We have found that the $\mf{sl}_3$ global one-point torus conformal block (\ref{1pt}), for $j=(r,s)$ and $j_1=(3a,0)$, is given by the following integral representation and relations
\begin{equation} \label{sl3block}
\begin{split}
 &   F(j_1, j,q)= \frac{1}{\mathcal{N} (r,s,a)}\int_{C_{\mathfrak{sl}_3}}  dZ \xi(  
 j^{*\omega}, j_1^w, j| Z,\vec{z}_1, Z \cdot q)\big{|}_{z_1 \to 1}, \\ &
 \mathcal{F}(j_1, j,q)_{\mf{sl}_3}= F(-j_1, -j,q),
 \end{split}
\end{equation}
where the final expression for $F(j_1, j,q)$ is given below by (\ref{sl3sec36}). $\mathcal{N} (r,s,a)$ is a normalization factor chosen properly so that the expansion in $q$ starts from 1, as in (\ref{sl31ptper}). $\xi$ is from (\ref{invafun}), $j^{*w}= (2-r,2-s)$\footnote{In $\mathfrak{sl}_2$, this change is analogous to $1-\Delta$ of the conformal dimension $\Delta$.}, $dZ=dwdxdy$, $C_{\mathfrak{sl}3}$ is a proper integration contour defined below in (\ref{sl3sec31}), and
\begin{equation} \label{sl3sec7}
    Z\cdot q= (q^2 w, q x,qy).
\end{equation}
In the remaining part of this section, we will justify (\ref{sl3block}) along the lines of the sphere case and compute the rhs of the first line of (\ref{sl3block}). To proceed, we redefine (\ref{1pt}) in terms of $\mf{sl}_3$ fields, namely
\begin{equation} \label{sl3block2}
  F (j_1,j,q) = \sum_{\substack{M \in \mathbb{I},\\|M|=|N|}}\sum_{N \in \mathbb{I}} (B_{j} ^{-1})^{M N} \bra{ j, M}  \Phi_{j_1^{w}}(\vec{z}_1,   \vec{\bar{z}}_1 ) q^{L_0}\ket{N, j}.
\end{equation}
The external and intermediate fields $ \Phi_{j_1^w} (\vec{z}_1, \vec{\bar{z}}_1), \Phi_{j}(Z, \bar{Z})$ are $\mathfrak{sl}_3$ fields that depend on isospin variables. The descendant states are given as follows  
\begin{equation} \label{sl3sec23}
    |N, j\rangle = W_{-2}^{n_3}W_{-1}^{n_2} L_{-1}^{n_1} \Phi_{j} (0,0) \ket{0}, \quad N = (n_1, n_2, n_3), \quad |N|=  n_1+n_2+2n_3.
\end{equation}
The operators $L_0,W_{-2}, L_{-1}, W_{-1}$ satisfy the algebra (\ref{eqn:W3.1}) and are given by a linear combination of differential operators (\ref{sl3sec3}). 

We use the conjectured holomorphic $\mathfrak{sl}_3$ operator product expansion 
\begin{equation} \label{sl3sec25}
\Phi_{ j} (Z , 0)\ket{0} = \sum_{N \in \mathbb{I}} a_{N}(j,Z) \ket{N, j} = \sum_{N \in \mathbb{I}} \beta(j,{N})  x^{n_1} y^{n_2} w^{n_3} \ket{N, j}.
\end{equation}
From this expansion, we have
\begin{equation} \label{factorsl31}
    \bra{j, M }  \Phi_{\tilde{j}} (Z, 0) \ket{0} = \delta_{j, \tilde{j}}\sum_{L \in \mathbb{I}} a_{L} (j,Z) B_{j}^{ M L}.
\end{equation}
 We insert $\mathbf{P}$ into (\ref{sl3block2}), obtaining
 \begin{equation}  \label{sl3sec26}
 \begin{split}
  & F (j_1,j,q) = \sum_{\substack{M \in \mathbb{I},\\ |M|=|N|}}\sum_{N \in \mathbb{I}} (B_{j} ^{-1})^{M N} \bra{ j, M} \mathbf{P}  \Phi_{j_1^w}(\vec{z}_1,   \vec{\bar{z}}_1 ) q^{L_0}\ket{N, j}= \\&  \sum_{\substack{M \in \mathbb{I},\\ |M|=|N|}}\sum_{N \in \mathbb{I}} (B_{j} ^{-1})^{M N} \bra{ j, M} \sum_{\tilde{j}} \int d^2 Z \Phi_{\tilde{j}}(Z, \bar{Z}) \ket{0} \bra{0} \tilde{\Phi}_{\tilde{j}^{*w}} (Z, \bar{Z})  \Phi_{j_1^w}(\vec{z}_1,   \vec{\bar{z}}_1 ) q^{L_0}\ket{N, j}.
\end{split}
\end{equation}
Using (\ref{factorsl31}) in the factor $\bra{j  , M}  \Phi_{\tilde{j}}(Z, \bar{Z})\ket{0}$, and $q^{L_0} \ket{N, j}= q^{\Delta (j)+|N|   } \ket{N, j}$ and after the cancelation of the matrices $B^{-1}$ and $B$, we obtain
\begin{equation} \label{sl3sec27}
     F (j_1,j,q) = \int d^2Z  \sum_{N \in \mathbb{I}} q^{\Delta (j)+|N|   }\beta(j, N) x^{n_1} y^{n_2} w^{n_3}  \bra{0} \tilde{\Phi}_{ j^{*w}} (Z, \bar{Z})  \Phi_{j_1^w}(\vec{z}_1,   \vec{\bar{z}}_1 ) \ket{N, j}.
\end{equation}
Since $|N|= n_1+n_2+2n_3$, we can write $q^{|N|}= q^{n_1+n_2+2n_3}$, and applying again (\ref{factorsl31}), we obtain
\begin{equation} \label{sl3sec28}
     F (j_1,j,q) = q^{\Delta (j) }\int d^2Z \bra{0} \tilde{\Phi}_{j^{*w}} (Z, \bar{Z})  \Phi_{j_1^w}(\vec{z_1},   \vec{\bar{z}}_1 ) \Phi_{j} (Z.q,0)\ket{0}.
\end{equation}
For the coordinate dependence of the three-point correlation function of $\mf{sl}_3$ fields, according to the above discussion, we use the formula
\begin{equation} \label{sl3sec29}
    \bra{0} \Phi_{j_1} (Z_1, \bar{Z}_1)  \Phi_{j_2}(Z_2,   \bar{Z}_2 ) \Phi_{j_3} (Z_3,\bar{Z}_3)\ket{0} =|\xi(j_1, j_2, j_3|Z_1, Z_2, Z_3)|^2,
\end{equation}
hence
\begin{equation} \label{sl3sec30}
     F (j_1,j,q) = q^{\Delta (j) }\int d^2Z \xi(j^{*w}, j_1^w, j|Z, \vec{z}_1, Z.q) \xi(j^{*w}, j_1^w, j|\bar{Z}, \vec{\bar{z}}_1, 0).
\end{equation}
Equation (\ref{sl3sec30}) is the direct analog to (\ref{integral1}). As explained in observation II of section \bref{1ptpw1}, to obtain the CB from (\ref{sl3sec30}), one needs to find the proper integration contour in order to simplify the above integral. Furthermore, after replacing $z_1$ in (\ref{sl3sec30}) by using (\ref{invafun}), we notice that it is possible to take out $z_1$ as an overall factor. Since $z_1$ does not play an important role in our discussion, we set $z_1=1$. Under this consideration, we found that this contour is
 \begin{equation} \label{sl3sec31}
      C_{\mathfrak{sl}_3}: \quad x \in \left(0,1/2\right), \quad w \in \left(x-1/2, 0\right), \quad y\in \left(\frac{w}{x}, \frac{w}{x-1/2} \right).
 \end{equation}
 Hence, (\ref{sl3sec30}) can be simply written as the first line of (\ref{sl3block}), where in the normalization coefficient $\mathcal{N} (r,s,a)$ of (\ref{sl3block}) we absorbed the factor $q^{\Delta(j)}$. Finally, we compute the first line of (\ref{sl3block}). Our final result precisely reproduces (\ref{sl31ptper}), namely, we found the relation given in the second line of (\ref{sl3block}).  
The expression we found for $F(j_1, j,q)$  from (\ref{sl3block}) is given by the integral 
\begin{equation}\label{sl3sec35}
\begin{split}
  &   F(j_1,j,q) =   \frac{1}{\mathcal{N} (r,s,a)} \int_{0}^{\frac{1}{2}}dx\int_{x-\frac{1}{2}} ^{0} dw  \int_{ \frac{w}{x}     }^{\frac{w}{x-\frac{1}{2}}} dy   \xi(  
 j^{*\omega}, j_1^w, j| Z,\vec{z}_1, Z \cdot q)\big{|}_{z_1 \to 1}\\&
=\frac{1}{\mathcal{N} (r,s,a)} \int_{0}^{\frac{1}{2}}dx\int_{x-\frac{1}{2}} ^{0} dw  \int_{ \frac{w}{x}     }^{\frac{w}{x-\frac{1}{2}}} dy \left(q^2 w+y (x-q x)-w\right)^{a+s-2} \\& \begin{split} & \bigg{(}-q^2w y+q^2 w z_1+q w y+q y^2 (q x-x)+q y z_1 \left(z_1-q x\right)-\frac{1}{2} q y z_1^2-w z_1+y z_1 \left(x-z_1\right)+ \\& \frac{y z_1^2}{2}\bigg{)}{}^{-a} \left(w+z_1 \left(z_1-x\right)-\frac{z_1^2}{2}\right){}^{-a+r+s-2} \left(q^2 w+z_1 \left(z_1-q x\right)-\frac{z_1^2}{2}\right){}^{-a-r-s+2} \times \\& \times \left(-q^2 w+q y (q x-x)+w\right)^{a-s} \bigg{|}_{z_1 \to 1}.
  \end{split} 
    \end{split}
\end{equation}
In appendix \bref{thesl3integral}, we explain how we compute this integral; here we present the final result

\begin{equation}  \label{sl3sec36}
\begin{split}
    & F(j_1,j,q) = \\& \frac{ (1-q)^{a-2} }{\tilde{\mathcal{N}}}  \sum _{i,k,l,m=0}^{\infty }  \sum _{j=0}^i  \sum_{n=0}^k  \frac{q^{i+j+k+l+m}   (-1)^{i+k+n} \Gamma (a+k) \Gamma (-a-k+1) C^{i}_{j} C^{k}_{n} }{\Gamma (k+1) \Gamma (i+r+s-1) \Gamma (j+n+r+s-1) \Gamma (-l+m+n+s)} \times \\&  C^{a+s-2}_{l} C^{a-s}_{m} \Gamma (a+j+n) C^{-a-r-s+2}_{i}  \Gamma (-a+j-k+l-m+r) \Gamma (a+k-l+m+n+s-1)\\& \Gamma (a+i-j+k-l+m+s-1) {}_3F_2(A,B,C;B_1,B_2;1),
\end{split} 
\end{equation}
where 
\begin{equation} \label{sl3sec37}
\begin{split}
  &  A=  -a+j-k+l-m+r, \quad B=  -a-k+l-m-s+2, \quad C=l-m-n-s+1, \\& 
  B_1=-a-i+j-k+l-m-s+2, \quad B_2=-a-k+l-m-n-s+2.
\end{split}
\end{equation}
and $_3F_2$ is the hypergeometric function and $C^i_j=\binom{i}{j}$, and 
\begin{equation} \label{sl3sec38}
    \tilde{\mathcal{N}}= \frac{ \Gamma (1-a) (\Gamma (a))^2 \Gamma (r-a) (\Gamma (a+s-1))^2 \, _2F_1(r-a,1-s,-a-s+2, 1)}{\Gamma (s) (\Gamma (r+s-1))^2}.
    \end{equation}
Equation (\ref{sl3sec36}) together with (\ref{sl3block}) are our main results. 
\section{Conclusions} \label{conclusions}
In this work, we have studied global conformal blocks using shadow formalism. Our study focused on $\mathfrak{sl}_2$ and $\mathfrak{sl}_3$ global CBs, which arise in the large central charge limit of Virasoro and $\mathcal{W}_3$ conformal field theories. In sections \bref{sphereCBPW} and \bref{1ptpw1}, we examined the representation of sphere and torus $\mathfrak{sl}_2$ global CBs through the shadow formalism. While our primary focus was on the torus topology, we also discussed the spherical case. For $\mathfrak{sl}_2$, our discussion built upon ideas from previous works, we introduced an operator (\ref{sl2gbs11}) that acts as a resolution of identity. This allows us to express the CBs in terms of conformal partial waves. We showed that the $\mathfrak{sl}_2$ global four-point sphere and one-point torus CBs can be computed using shadow formalism. The construction is quite universal and can be easily generalized to higher multi-point global torus or spherical conformal blocks.

In section \bref{geralsl3shf}, we investigated $\mathcal{W}_3$ CFT and generalized the shadow formalism to this case; this enables the computation of $\mathfrak{sl}_3$ global CBs. The formalism was then applied to both spherical and toroidal CBs in sections \bref{sl34ptsphereCB} and \bref{sl3globalonepointtorus}. We verified that this approach yields the established expression (\ref{sl34ptsphere4}) for the $\mf{sl}_3$ global four-point sphere CBs. In Section \bref{sl3globalonepointtorus}, we obtained an integral representation for the $\mathfrak{sl}_3$ global one-point torus CB (\ref{sl3block}), resulting in the explicit expression (\ref{sl3sec36}). This expression is in complete agreement with the known perturbative expressions (\ref{sl31ptper}). The expression we obtained for $\mathfrak{sl}_3$ global one-point torus CB represents the main result of the present study.

It is interesting to investigate the possibility of constructing $\mathfrak{sl}_3$ global higher multi-point torus CBs involving different OPE channels. In the $\mathfrak{sl}_2$ case, $(n+2)$-point sphere CBs correlate with $n$-point torus CBs. It would be interesting to find out whether a similar connection exists in the case of $\mathfrak{sl}_3$. Another direction for generalization is the development of the shadow formalism for general $\mathcal{W}_N$ two-dimensional CFTs or supersymmetric ones. As demonstrated in the $\mathfrak{sl}_2$ and $\mathfrak{sl}_3$ cases, such generalization requires the analysis of invariant functions related to the corresponding algebras.

\appendix
\section{Conformal Integrals} \label{app.CI}
We consider  the conformal integrals
\begin{equation}
    I_n= \frac{1}{\pi}  \int d^2x f_n(z) \bar{f}_n (\bar{z}),\quad f_n(z)=\prod _{i=1}^n  \frac{1}{(z-z_i)^{h_i}}, \quad  \bar{f}_n(\bar{z})=\prod _{i=1}^n  \frac{1}{(\bar{z}-\bar{z}_i)^{\bar{h_i}}},
\end{equation}
where
\begin{equation}
    \sum_{i=1}^n h_i=  \sum_{i=1}^n \bar{h}_i=2,  \quad h_i-\bar{h}_i=\mathbf{Z}.
\end{equation}
For $n=2$, the result is known to be
\begin{equation} \label{I2formula}
I_2 = K_{12}(-1)^{h_1-\bar{h}_1}  \pi \delta^2 (x_1-x_2),
\end{equation}
where 
\begin{equation}
    K_{12} = \frac{\Gamma(1-h_1) \Gamma(1-h_2)}{\Gamma(\bar{h}_1)   \Gamma( \bar{h}_2)   }.
\end{equation}
For $n=3$, it is known that
\begin{equation} \label{cint3}
    I_3= K_{123} z_{12}^{h_3-1}z_{23}^{h_1-1}z_{31}^{h_2-1}    \bar{z}_{12}^{  \bar{h}_3-1}\bar{z}_{23}^{\bar{h}_1-1}\bar{z}_{31}^{\bar{h}_2-1},
\end{equation}
where
\begin{equation} \label{K123}
    K_{123}(h_1, h_2, h_3) = \frac{\Gamma(1-h_1) \Gamma(1-h_2) \Gamma(1-h_3)   }{  \Gamma(\bar{h}_1)  \Gamma(\bar{h}_2)  \Gamma(\bar{h}_3)}.
\end{equation}
\section{$\mathfrak{sl}_3$ Integral} \label{thesl3integral}
Below, we explain some technical details of the computation of the integral (\ref{sl3sec35}). Let us denote that integral by $I_{0}$,
 \begin{equation}
\begin{split}
  &    I_0 =    \int_{0}^{\frac{1}{2}}dx\int_{x-\frac{1}{2}} ^{0} dw  \int_{ \frac{w}{x}     }^{\frac{w}{x-\frac{1}{2}}} dy  \xi(  
 j^{*\omega}, j_1^w, j| Z,\vec{z}_1, Z \cdot q)\bigg{|}_{z_1 \to 1}\\&
=\int_{0}^{\frac{1}{2}}dx\int_{x-\frac{1}{2}} ^{0} dw  \int_{ \frac{w}{x}     }^{\frac{w}{x-\frac{1}{2}}} dy \left(q^2 w+y (x-q x)-w\right)^{a+s-2} \left(-q^2 w+q y (q x-x)+w\right)^{a-s} \\& \begin{split} & \bigg{(}-q^2w y+q^2 w z_1+q w y+q y^2 (q x-x)+q y z_1 \left(z_1-q x\right)-\frac{1}{2} q y z_1^2-w z_1+y z_1 \left(x-z_1\right)+ \\& \frac{y z_1^2}{2}\bigg{)}{}^{-a} \left(w+z_1 \left(z_1-x\right)-\frac{z_1^2}{2}\right){}^{-a+r+s-2} \left(q^2 w+z_1 \left(z_1-q x\right)-\frac{z_1^2}{2}\right){}^{-a-r-s+2} \bigg{|}_{z_1 \to 1}\;.
  \end{split} 
    \end{split}
\end{equation}
 After some simplifications, we obtain
  \begin{equation} \label{asl3int1}
\begin{split}
    I_0 &=    (-1)^{-s}8 ^a  \int_{0}^{\frac{1}{2}}dx\int_{x-\frac{1}{2}} ^{0} dw  \int_{ \frac{w}{x}     }^{\frac{w}{x-\frac{1}{2}}}  dy    \frac{(1-q)^{a-2} (2 w-2 x+1)^{-a+r+s-2} }{(2 (q (y-1)-1) (x y-w)+y)^a}  \times \\ &
\left(2 q^2 w-2 q x+1\right)^{-a-r-s+2} (q w+w-x y)^{a+s-2} (q w-q x y+w)^{a-s} .
\end{split}
\end{equation}
Expanding in $q$ we obtain
\begin{equation} \label{asl3int2}
\begin{split}
    & I_0 =    (-1)^{-s}8 ^a    \int_{0}^{\frac{1}{2}}dx\int_{x-\frac{1}{2}} ^{0} dw  \int_{ \frac{w}{x}     }^{\frac{w}{x-\frac{1}{2}}}  dy \sum_{j=0}^i \sum_{i,k,l,m=0}\frac{\left((1-q)^{a-2} (2 w-2 x+1)^{-a+r+s-2}\right)}{\Gamma (a) \Gamma (k+1) \Gamma (m+1) \Gamma (s-a)}\times \\& \bigg{(}2^{i+k} (y-1)^k \Gamma (a+k) C^{i}_{j} (-1)^{i-j-m} x^{i-j} C^{a+s-2}_{l}  \Gamma (-a+m+s) C^{-a-r-s+2}_{i} \\& (2 w-2 x y+y)^{-a-k} w^{a+j+l-m-s} q^{i+j+k+l+m}  (w-x y)^{a+k-l+m+s-2}  \bigg{)}  .
\end{split}
\end{equation}
To integrate over $y$, we expand the factor $(y-a)^k$ in $y$. This expansion adds an extra sum from $0$ to $k$ present in the final result (\ref{sl3sec36}). The result of the integration over $y$ is given by a function of $w,x$ which contains a hypergeometric function ${}_2F_1$
with the argument $1 - \frac{1}{2x}$. Then, we perform the integration over $w$. The integration over $w$ has the form
\begin{equation} \label{asl3int3}
    \int_{x-\frac{1}{2}}^0 w^{\alpha } (2 w-2 x+1)^{-2 - a + r +s } \, dw,
\end{equation}
where $\alpha$ is some power. The result of (\ref{asl3int3}) is given by a ratio of gammas functions. Finally, to compute the integration over $x$, we expand the above-mentioned hypergeometric function in $1 - \frac{1}{2x}$ (the sum from this expansion was already considered in (\ref{sl3sec36}) and  produces the factor ${}_3F_2$). Thus, the integration over $x$ has the form

\begin{equation} \label{asl3int4}
    \int_{0}^{\frac{1}{2}} x^ {\alpha } (1-\frac{1}{2x})^{\beta } \, dx,
\end{equation}
where $\alpha, \beta$ are again some powers. After integrating over $y,w,x$, the final result is given in (\ref{sl3sec36}). 

\section{$\mf{sl}_3$ Fields} \label{apsl3fields}
In this appendix, we briefly describe the $\mf{sl}_3$ fields denoted above as $\Phi_j (Z, \bar{Z},g)$. For a more detailed description, see \cite{Fateev:2011qa}. The fields $\Phi_{j}(Z, \bar{Z}, g)$, for $g \in SL_3$, are basis functions on $SL_3$ defined as
\begin{equation} \label{apsl3f1}
    \Phi_{j}(Z, \bar{Z},g) = \left(u_Z P g^{-1 T}Pu_{\bar{Z}}^T \right)^{-r} \left( v_Z g v_{\bar{Z}}^T\right)^{-s},
\end{equation}
where 
\begin{equation}
    u_Z = (w,-x,1), \quad v_{Z}= (xy-w, -y,1),\quad P=  
    \begin{pmatrix}   0 & 0 &1\\ 0 & -1&0\\ 1&0 &0 \end{pmatrix}.
\end{equation}
The $\mf{sl}_3$ generators $t^a$ can be represented as matrices
\begin{equation}
    h^1= \begin{pmatrix}
         1 & 0 &0\\ 0 & -1&0\\ 0&0 &0
    \end{pmatrix}, \quad     h^2= \begin{pmatrix}
         0 & 0 &0\\ 0 & 1&0\\ 0&0 &-1
    \end{pmatrix}, \quad 
\end{equation}
\begin{equation}
    e^1= \begin{pmatrix}
         0 & 1 &0\\ 0 & 0&0\\ 0&0 &0
    \end{pmatrix}, \quad     e^2= \begin{pmatrix}
         0 & 0 &0\\ 0 & 0&1\\ 0&0 &0
    \end{pmatrix}, \quad 
    e^3= \begin{pmatrix}
         0 & 0 &1\\ 0 & 0&0\\ 0&0 &0
    \end{pmatrix}, 
\end{equation}
\begin{equation}
    f^1= \begin{pmatrix}
         0 & 0 &0\\ 1 & 0&0\\ 0&0 &0
    \end{pmatrix}, \quad     f^2= \begin{pmatrix}
         0 & 0 &0\\ 0 & 0&0\\ 0&1 &0
    \end{pmatrix}, \quad 
    f^3= \begin{pmatrix}
         0 & 0 &0\\ 0 & 0&0\\ 1&0 &0
    \end{pmatrix}.
\end{equation}
One can show that the fields (\ref{apsl3f1}) under $\mf{sl}_3$ transformation of $g$ transform according to (\ref{sl3trans}).
\bibliographystyle{JHEP} 
\bibliography{refs} 
\end{document}